# After 100 Years, Can We Finally Crack Post's Problem of Tag? A Story of Computational Irreducibility, and More

Stephen Wolfram*

*Empirical, theoretical and historical aspects of Post's "problem of tag" from 1921 are explored. Evidence of strong computational irreducibility is found. Despite their deterministic origin, the lengths of successive sequences generated seem to closely approximate random walks. All $10^{25}$ smallest initial conditions are found to eventually halt, although sometimes in $> 6 \times 10^{11}$ steps. Implications of the Principle of Computational Equivalence are discussed, along with examples of identifiable computational capabilities of tag systems. Various minimal examples of complex behavior are found, including a less-biased analog of the 3n+1 Collatz problem. There is also discussion of the history of Emil Post and of tag systems in the context of ideas about the foundations of mathematics and computation.*

## "[Despite] Considerable Effort… [It Proved] Intractable"

In the early years of the twentieth century it looked as if—if only the right approach could be found—all of mathematics might somehow systematically be solved. In 1910 Whitehead and Russell had published their monumental Principia Mathematica showing (rather awkwardly) how all sorts of mathematics could be represented in terms of logic. But Emil Post wanted to go further. In what seems now like a rather modern idea (with certain similarities to the core structure of the Wolfram Language, and very much like the string multiway systems in our Physics Project), he wanted to represent the logic expressions of Principia Mathematica as strings of characters, and then have possible operations correspond to transformations on these strings.

In the summer of 1920 it was all going rather well, and Emil Post as a freshly minted math PhD from Columbia arrived in Princeton to take up a prestigious fellowship. But there was one final problem. Having converted everything to string transformations, Post needed to have a theory of what such transformations could do.

He progressively simplified things, until he reached what he called the problem of "tag". Take a string of 0s and 1s. Drop its first $v$ elements. Look at the first dropped element. If it's a 0 add a certain block of elements at the end of the string, and if it's a 1 add another block. Post solved several cases of this problem.

But then he came across the one he described as 0→00, 1→1101 with $v$ = 3. Here's an example of its behavior:

101
1101
11101
011101
10100
001101
10100
001101
10100
001101
10100

After a few steps it just ends up in a simple loop, alternating forever between two strings. Here's another example, starting now from a different string:

10010
101101
1011101
11011101
111011101
0111011101
101110100
1101001101
10011011101
110111011101
1110111011101
01110111011101
1011101110100
11011101001101
111010011011101
0100110111011101
011011101110100
01110111010000
1011101000000
11010000001101
100000011011101
0000110111011101
011011101110100
01110111010000
1011101000000
11010000001101
100000011011101
0000110111011101
011011101110100
01110111010000
1011101000000

Again this ends up in a loop, now involving 6 possible strings.

But what happens in general? To Post, solving this problem was a seemingly simple stepping stone to his program of solving all of mathematics. And he began on it in the early summer of 1921, no doubt expecting that such a simple-to-state problem would have a correspondingly simple solution.

But rather than finding a simple solution, he instead discovered that he could make little real progress. And after months of work he finally decided that the problem was in fact, as he later said, "hopeless"—and as a result, he concluded, so was his whole approach to "solving mathematics".

What had happened? Well, Post had seen a glimpse of a completely unanticipated but fundamental feature of what we now call computation. A decade later what was going on became a little clearer when Kurt Gödel discovered Gödel's theorem and undecidability. (As Post later put it: "I would have discovered Gödel's theorem in 1921—if I had been Gödel.") Then as the years went by, and Turing machines and other kinds of computational systems were introduced, tag systems began to seem more about computation than about mathematics, and in 1961 Marvin Minsky proved that in fact a suitably constructed tag system could be made to do any computation that any Turing machine could do.

But what about Post's particular, very simple tag system? It still seemed very surprising that something so simple could behave in such complicated ways. But sixty years after Post's work, when I started to systematically explore the computational universe of simple programs, it began to seem a lot less surprising. For—as my Principle of Computational Equivalence implies—throughout the computational universe, above some very low threshold, even in systems with very simple rules, I was seeing the phenomenon of computational irreducibility, and great complexity of behavior.

But now a century has passed since Emil Post battled with his tag system. So armed with all our discoveries—and all our modern tools and technology—what can we now say about it? Can we finally crack Post's problem of tag? Or—simple as it is—will it use the force of computational irreducibility to resist all our efforts?

This is the story of my recent efforts to wage my own battle against Post's tag system.

## The Basic Setup

The Wolfram Language can be seen in part as a descendent of Post's idea of representing everything in terms of transformation rules (though for symbolic expressions rather than strings). So it's no surprise that Post's problem of tag is very simple to state in the Wolfram Language:

```
NestList[Replace[{
        {0, _, _, s___} → {s, 0, 0},
        {1, _, _, s___} → {s, 1, 1, 0, 1}
    }], {1, 0, 0, 1, 0}, 10] // Column
```

{1, 0, 0, 1, 0}
{1, 0, 1, 1, 0, 1}
{1, 0, 1, 1, 1, 0, 1}
{1, 1, 0, 1, 1, 1, 0, 1}
{1, 1, 1, 0, 1, 1, 1, 0, 1}
{0, 1, 1, 1, 0, 1, 1, 1, 0, 1}
{1, 0, 1, 1, 1, 0, 1, 0, 0}
{1, 1, 0, 1, 0, 0, 1, 1, 0, 1}
{1, 0, 0, 1, 1, 0, 1, 1, 1, 0, 1}
{1, 1, 0, 1, 1, 1, 0, 1, 1, 1, 0, 1}
{1, 1, 1, 0, 1, 1, 1, 0, 1, 1, 1, 0, 1}

Given the initial string, the complete behavior is always determined. But what can happen? In the examples above, what we saw is that after some "transient" the system falls into a cycle which repeats forever.

Here's a plot for all possible initial strings up to length 7. In each case there's a transient and a cycle, with lengths shown in the plot (with cycle length stacked on top of transient length):

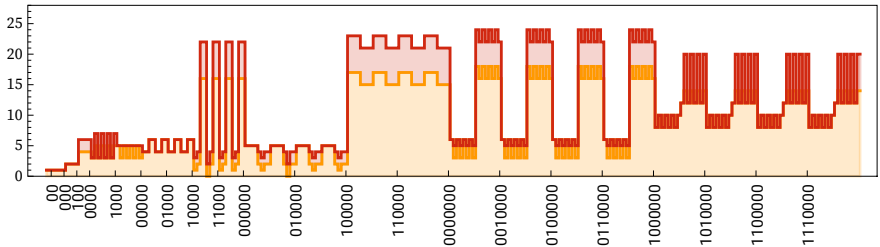

(Note that if the system reaches 00—or another string with less than 3 characters—one can either say that it has a cycle of length 1, or that it stops completely, effectively with a cycle of length 0.) For initial strings up to length 7, the nontrivial cycles observed are of lengths 2 and 6.

Starting from 10010 as above, we can show the behavior directly—or we can try to compensate for the removal of elements from the front at each step by rotating at each step:

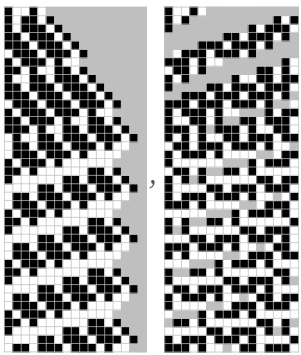

We can also show only successive "generations" in which the rule has effectively "gone through the whole string":

Let's continue to longer initial sequences. Here are the lengths of transients and cycles for initial sequences up to length 12:

All the cycles are quite short—in fact they're all of lengths 0, 2, 4, 6 or 10. And for initial strings up to length 11, the transients (which we can think of as "halting times") are at most of length 28. But at length 12 the string 100100100000 suddenly gives a transient of length 419, before finally evolving to the string 00.

Here's a plot of the sequence of lengths of intermediate strings produced in this case (the maximum length is 56):

And, by the way, this gives an indication of why Post called this the "problem of tag" (at the suggestion of his colleague Bennington Gill). Elements keep on getting removed from the "head" of the string, and added to its "tail". But will the head catch up with the tail? When it does, it's like someone winning a game of tag, by being able to "reach the last person".

Here's a picture of the detailed behavior in the case above:

And here's the "generational" plot, now flipped around to go from left to right:

By the way, we can represent the complete history of the tag system just by concatenating the original string with all the blocks of elements that are added to it, never removing blocks of elements at the beginning. In this case this is the length-1260 string we get:

```
100100100000110111011101001101110100110100110111011101110111011101110100110111011101001101110111011101110100110100110110111101...
110100110111011101110111011101101001101001101110011011101110100110111011101001101001101001101110111010000011011101110111101...
110111011101110110100110100000110111010011011101110100110111011101001101001101000000110100110111011101000000110111011...
1101110111011101110100000000011011101110100000001101110111010011011101110100000000110111010011010000110111010001010100...
110111011101000000000110100110100000011010011010011011101110100000001101110111010000000000001101110110111010000...
110111010011011101110100000000110111010011010000011010011011101110111010000000110100110100001101110111101001101000...
0000000000001101110100110111011101110111010000000001101110111011101110100110111011101000000011011101001101110111...
01000001101110111010000011011101000000110111011101000000110111011101001101110111010000001101110111010000011011101111...
01110111010011010001101110100110100001101110100110110111010000000011011101000000110111010000001101110111010000111...
0111010000000000110100001101110100110100000011010000000000011011101000000000110101000000011011101000000011011101100...
0011011101000000000110101000000110100000000000011010000000000110100000000000000000000
```

Plotting the "walk" obtained by going up at each 1 and down at each 0 we get (and not surprisingly, this is basically the same curve as the sequence of total string lengths above):

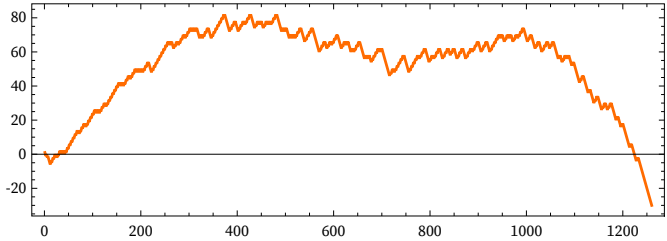

How "random" is the sequence of 0s and 1s? There are a total of 615 1s and 645 0s in the whole sequence—so roughly equal. For length-2 blocks, there are only about 80% as many 01s and 10s as 00s and 11s. For length-3 blocks, the disparities are larger, with only 30% as many 001 blocks occurring as 000 blocks.

And then at length 4, there is something new: none of the blocks

{0101, 1100, 1111}

ever appear at all, and 0010 appears only twice, both at the beginning of the sequence. Looking at the rule, it's easy to see why, for example, 1111 can never occur—because no sequence of the 00s and 1101s inserted by the rule can ever produce it. (We'll discuss block occurrences more below.)

OK, so we've found some fairly complicated behavior even with initial strings of length 12. But what about longer strings? What can happen with them? Before exploring this, it's useful to look in a little more detail at the structure of the underlying problem.

## The Space of Possible States

To find out what can happen in our tag system, we've enumerated all possible initial strings up to certain lengths. But it turns out that there's a lot of redundancy in this—as our plots of "halting times" above might suggest. And the reason is that the way the tag system operates, only every third element in the initial string actually ever matters. As far as the rule is concerned we can just fill in _ for the other elements:

```
0__1__1__1__1__
1__1__1__1__00
1__1__1__001101
1__1__0011011101
1__00110111011101
0011011101110111101
10111011101110100
110111011101001101
11101110100011011101
01110100110111011101
1010011011101110100
```

The _'s will steadily be "eaten up", and whether they were originally filled in with 0s or 1s will never matter. So given this, we don't lose any information by using a compressed representation of the strings, in which we specify only every third element:

| | |
|---|---|
| **0**01**1**10**1**10**1**00**1**00 | **01111**$_0$ |
| **1**10**1**10**1**00**1**00**0**0 | **11110**$_2$ |
| **1**10**1**00**1**00**0**01**1**01 | **11101**$_0$ |
| **1**00**1**00**0**01**1**01**1**101 | **110111**$_1$ |
| **1**00**0**01**1**01**1**10**1**1**1**01 | **101110**$_2$ |
| **0**01**1**01**1**10**1**11**0**11**1**01 | **011101**$_0$ |
| **1**01**1**10**1**11**0**11**1**0**1**00 | **111010**$_2$ |
| **1**10**1**11**0**11**1**01**0**01**1**01 | **110101**$_0$ |
| **1**11**0**11**1**0**1**00**1**10**1**1**1**01 | **1010111**$_1$ |
| **0**11**1**01**0**01**1**01**1**101**1**101 | **0101110**$_2$ |
| **1**01**0**01**1**01**1**10**1**1**1**0**1**00 | **1011100**$_1$ |

But actually this isn't quite enough. We also need to say the "phase" of the end of the string: the number of trailing elements after the last block of 3 elements (i.e. the length of the original string mod 3).

So now we can start enumerating non-redundant possible initial strings, specifying them in the compressed representation:

| | | | |
|---|---|---|---|
| 000$_0$ | 010$_0$ | 100$_0$ | 110$_0$ |
| 000$_1$ | 010$_1$ | 100$_1$ | 110$_1$ |
| 000$_2$ | 010$_2$ | 100$_2$ | 110$_2$ |
| 001$_0$ | 011$_0$ | 101$_0$ | 111$_0$ |
| 001$_1$ | 011$_1$ | 101$_1$ | 111$_1$ |
| 001$_2$ | 011$_2$ | 101$_2$ | 111$_2$ |

Given a string in compressed form, we can explicitly compute its evolution. The effective rules are a little more complicated than for the underlying uncompressed string, but for example the following will apply one step of evolution to any compressed string (represented in the form {phase, elements}):

Replace[
 {{0, {0, s___}} → {2, {s, 0}}, {0, {1, s___}} → {1, {s, 1, 1}},
  {1, {0, s___}} → {0, {s}}, {1, {1, s___}} → {2, {s, 0}},
  {2, {0, s___}} → {1, {s, 0}}, {2, {1, s___}} → {0, {s, 1}}}]

Can we reconstruct an uncompressed string from a compressed one? Well, no, not uniquely. Because the "intermediate" elements that will be ignored by the rule aren't specified in the compressed form. Given, say, the compressed string 10:$_2$ we know the uncompressed string must be of the form 1__0_ but the _'s aren't determined. However, if we actually run the rule, we get

```
1__0_
0_1101
10100
001101
```

so that the blanks in effect quickly resolve. (By the way, given a compressed string s:0 the uncompressed one is $\bar{s}$ _ _, for s:1 it is just $\bar{s}$, and for s:2 it is $\bar{s}$ _, with the uncompressed string length mod 3 being equal to the phase.)

So taking all compressed strings up to length 4 here is the sequence of transient and cycle lengths obtained:

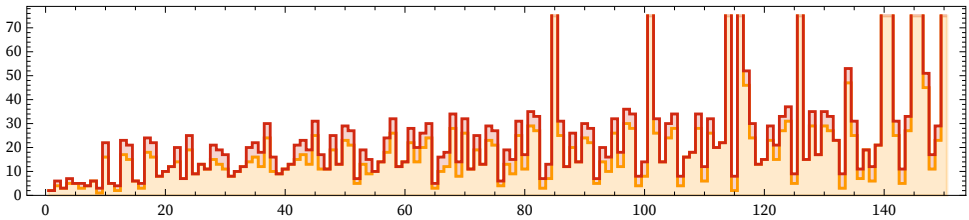

The first case that is cut off in the plot has halting time 419; it corresponds to the compressed string 1110:0.

We can think of compressed strings as corresponding to possible non-redundant "states" of the tag system. And then we can represent the global evolution of the system by constructing a state transition graph that connects each state to its successor in the evolution. Here is the result starting from distinct length-3 strings (here shown in uncompressed form; the size of each node reflects the length of the string):

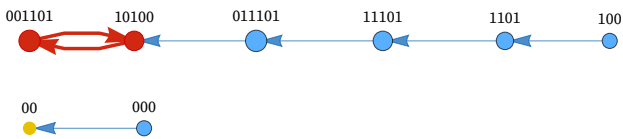

There is a length-2 cycle, indicated in red, and also a "terminating state" indicated in yellow. Here's the state transition graph starting with all length-1 compressed strings (i.e. non-redundant uncompressed strings with lengths between 3 and 5)—with nodes now labeled just with the (uncompressed) length of the string that they represent:

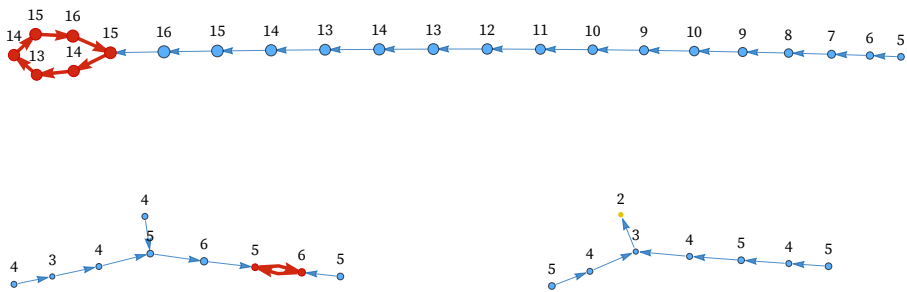

We see the same length-2 cycle and terminating state as we saw before. But now there is also a length-6 cycle. The original "feeder" for this length-6 cycle is the string 10010 (compressed: 11:2), which takes 16 steps to reach the cycle.

Here are the corresponding results for compressed initial strings up to successively greater lengths n, with the lengths of cycles labeled:

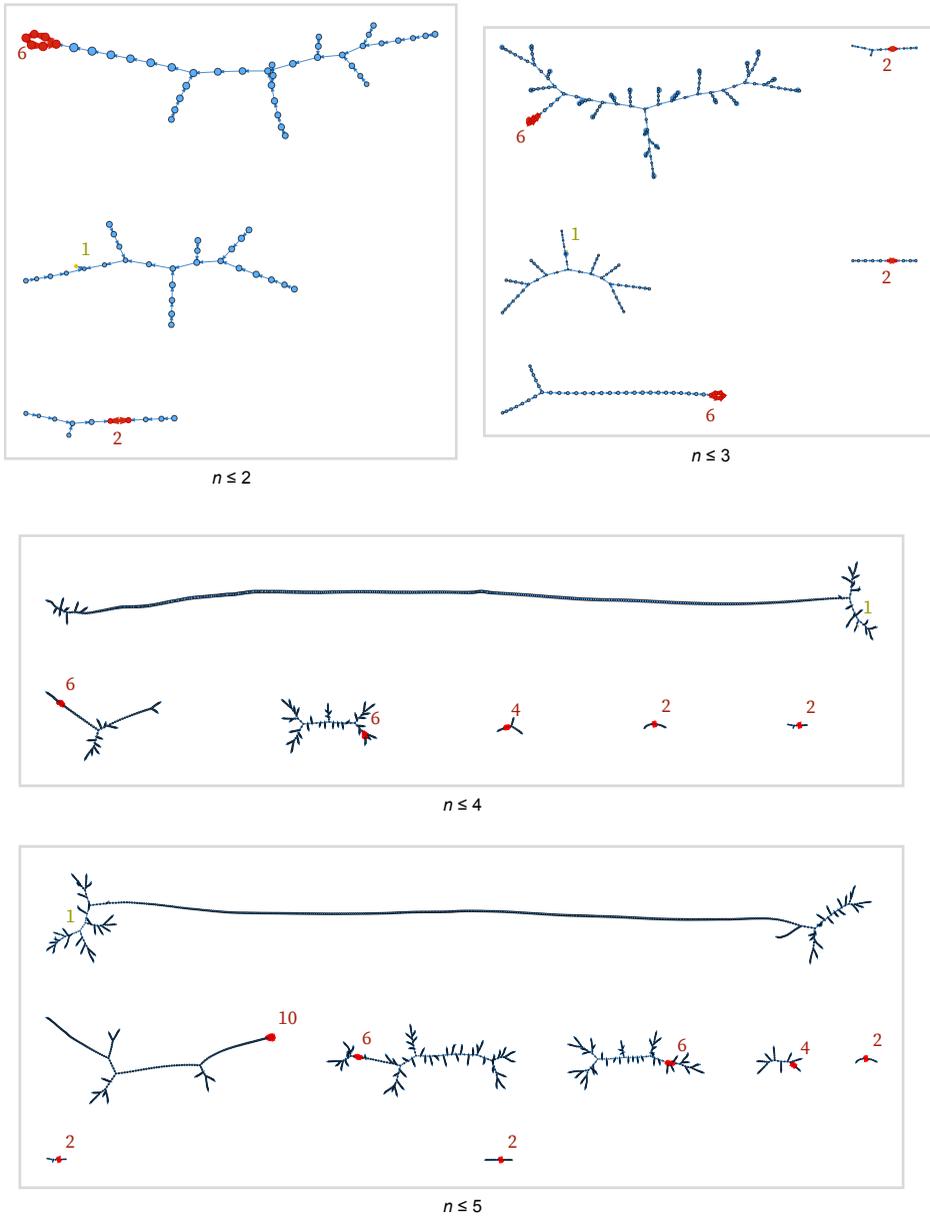

n ≤ 2

n ≤ 3

n ≤ 4

n ≤ 5

A notable feature of these graphs is that at compressed length 4, a long "highway" appears that goes for about 400 steps. The highway basically represents the long transient first seen for the initial string 11:2. There is one "on-ramp" for this string, but then there is also a tree of other states that enter the same highway.

Why is there a "highway" in the first place? Basically because the length-419 transient involves strings that are long compared to any we are starting from—so nothing can feed into it, and it basically just has to "work itself through" until it reaches whatever cycle it ends up in.

When we allow initial strings with compressed length up to 6 a new highway appears, dwarfing the previous one (by the way, most of the wiggliness we see is an artifact of the graph layout):

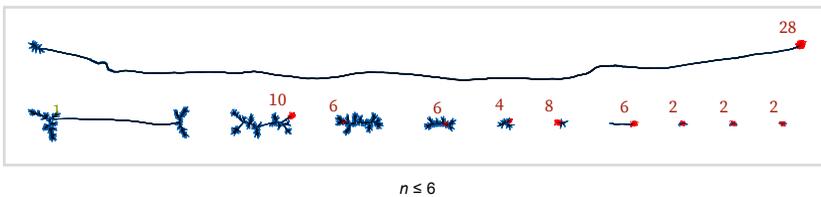

$n \leq 6$

The first initial state to reach this highway is 111010:0 (uncompressed: 100100100000100000)—which after 2141 steps evolves to a cycle of length 28. Here are the lengths of the intermediate strings along this highway (note the cycle at the end):

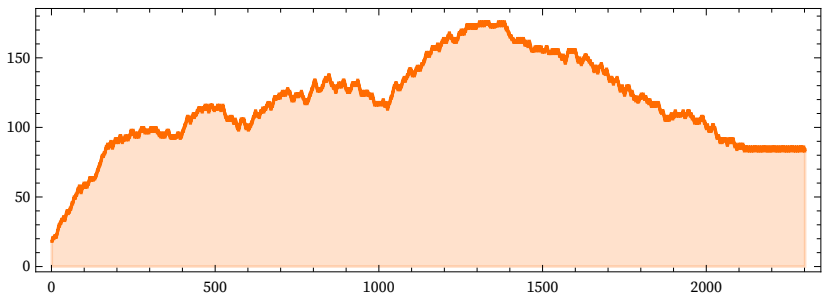

And here are the "generational states" reached (note that looking only at generations makes the final 28-cycle show up as a 1-cycle):

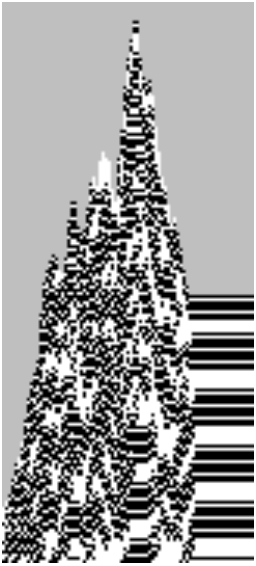

Or looking at "compressed strings" (i.e. including only every third element of each string):

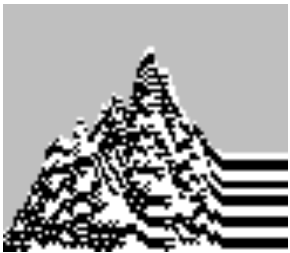

If we consider all initial strings up to compressed length 6, we get the following transient+cycle lengths:

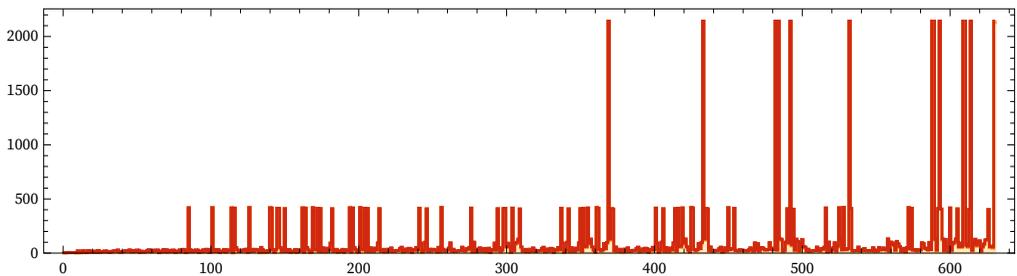

And what we see is that there are particular lengths of transients—corresponding to the highways in the state transition graph above—to which certain strings evolve. If we plot the distribution of halting (i.e. transient) times for all the strings we find, then, as expected, it peaks around the lengths of the main highways:

So given a particular "on-ramp to a highway"—or, for that matter, a state on a cycle—what states will evolve to it? In general there'll be a tree of states in the state transition graph that are the "predecessors" of a given state—in effect forming its "basin of attraction".

For any particular string the rule gives a unique successor. But we can also imagine "running the rule backwards". And if we do this, it turns out that any given compressed string can have 0, 1 or 2 immediate predecessors. For example, 000:0 has the unique predecessor 0000:1. But 001:0 has both 0001:1 and 100:2 as predecessors. And for example 001:1 has no predecessors. (For uncompressed strings, there are always either 0 or 4 immediate predecessors.)

Any state that has no predecessors can occur only as the initial string; it can never be generated in the evolution. (There are similar results for substrings, as we'll discuss later.)

And if we start from a state that does have at least one predecessor, we can in general construct a whole tree of "successively further back" predecessors. Here, for example, is the 10-step tree for 000:2:

Here it is after 30 steps, in two different renderings:

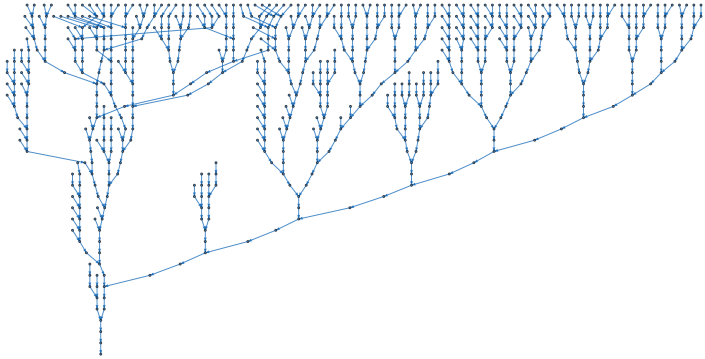

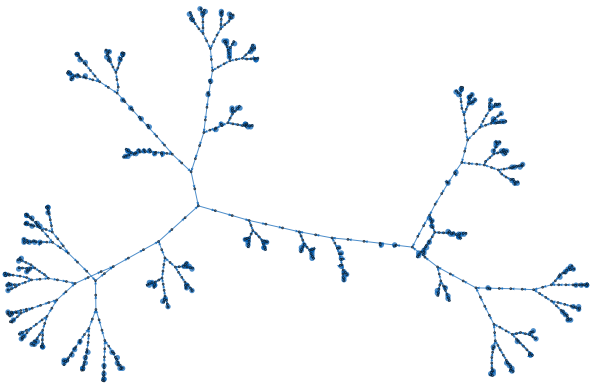

If we continue this particular tree we'll basically get a state transition graph for all states that eventually terminate. Not surprisingly, there's considerable complexity in this tree—though the number of states after t steps does grow roughly exponentially (apparently like $\approx 1.12^t$):

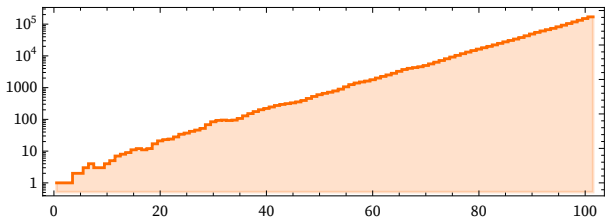

By the way, there are plenty of states that have finite predecessor trees. For example 1100:0 yields a tree which grows only for 21 steps, then stops:

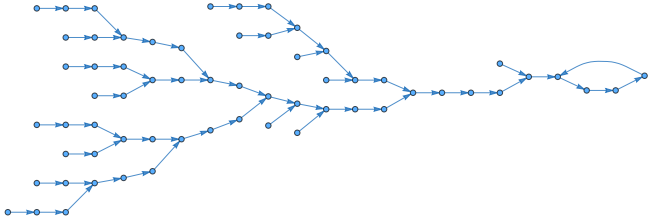

## The Cycle Structure

At least in all the cases we've seen so far, our tag system always evolves to a cycle (or terminates in a trivial state). But what cycles are possible? In effect any cycle state S must be a solution to a "tag eigenvalue equation" of the form $T^p S = S$ for some p, where T is the "tag evolution operator".

Starting with compressed strings of length 1, only one cycle can ever be reached:

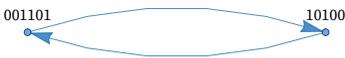

Starting with compressed strings of length 2 a 6-cycle appears (here shown labeled respectively with uncompressed and with compressed strings):

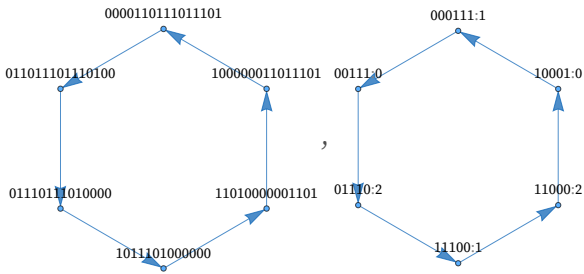

No new cycles appear until one has initial strings of compressed length 4, but then one gets (where now the states are labeled with their uncompressed lengths):

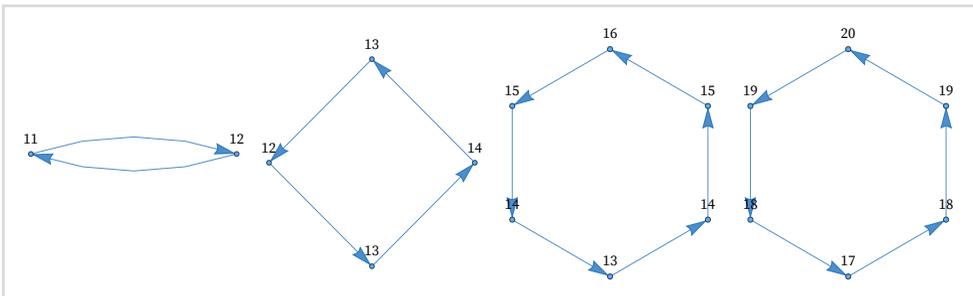

The actual cycles are as follows

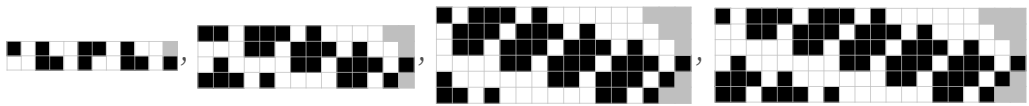

while the ones from length-5 initial strings are:

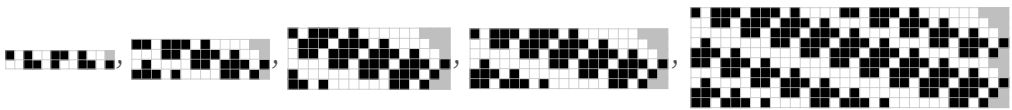

What larger cycles can occur? It is fairly easy to see that a compressed string consisting of any sequence of the blocks 01 and 1100 will yield a state on a cycle. To find out about uncompressed strings on cycles, we can just apply the rule 0→00, 1→1101, with the result that we conclude that any sequence of the length-6 and length-12 blocks 001101 and 110111010000 will give a state on a cycle.

If we plot the periods of cycles against the lengths of their "seed" strings, we get:

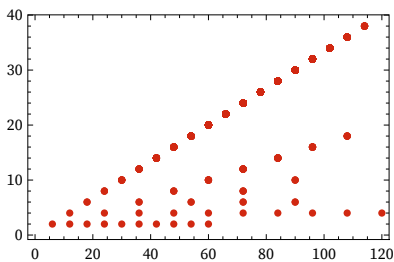

If we generate cycles from sequences of, say, b of our 01, 1100 blocks, how many of the cycles we get will be distinct? Here are the periods of the distinct cycles for successive b:

| 1 | {2, 4} |
|---|---|
| 2 | {2, 4, 6} |
| 3 | {2, 4, 8, 10} |
| 4 | {2, 4, 6, 10, 12, 14} |
| 5 | {2, 4, 12, 14, 14, 16, 16, 18} |
| 6 | {2, 4, 6, 8, 10, 14, 16, 16, 18, 18, 18, 20, 20, 22} |

The total number of cycles turns out to be:

DivisorSum[n, k ⟼ EulerPhi[k] $2^{n/k}$]/n

{2, 3, 4, 6, 8, 14, 20, 36, 60, 108, 188, 352, 632, 1182, 2192}

We can also ask an inverse question: of all $2^n$ (uncompressed) strings of length n, how many of them lie on cycles of the kind we have identified? The answer is the same as the number of distinct "cyclic necklaces" with n beads, each 0 or 1, with no pair of 0s adjacent:

DivisorSum[n, k ⟼ EulerPhi[n/k] × LucasL[k]]/n

{1, 2, 2, 3, 3, 5, 5, 8, 10, 15, 19, 31, 41, 64, 94, 143, 211, 329, 493, 766}

Asymptotically this is about $\phi^{n/2}$—implying that of all $2^n$ strings of length n, only a fraction ≈ $0.636^n$ of them will be on cycles, so that for large n the overwhelming majority of strings will not be on cycles, at least of this kind.

But are there other kinds of cycles? It turns out there are, though they do not seem to be common or plentiful. One family—always of period 6—are seeded by compressed strings of the form (00 111 (000 111))$^m$ (with uncompressed length 16 + 18m):

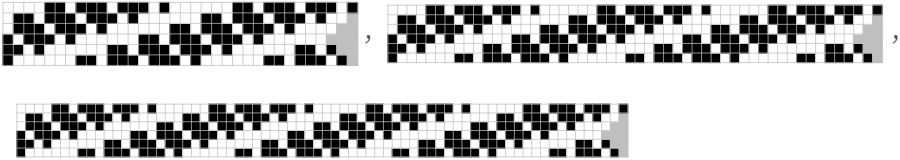

But there are other cases too. The first example appears with initial compressed strings of length 9. The length-13 compressed string 0011111110100 (with uncompressed length 39) yields the period-40 cycle (with uncompressed string lengths between 37 and 44):

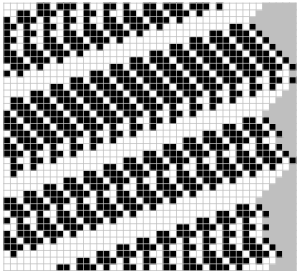

The next example occurs with an initial compressed string of length 15, and a compressed "seed" of length 24—and has period 282:

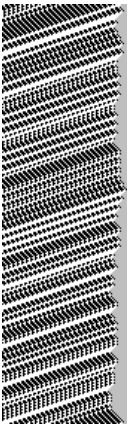

And I've found one more example (that arises from an initial compressed string of length 18) and has period 66:

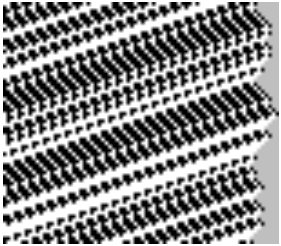

If we look at these cycles in "generational" terms, they are of lengths 3, 11 and 14, respectively (note that the second two pictures above start with "incomplete generations"):

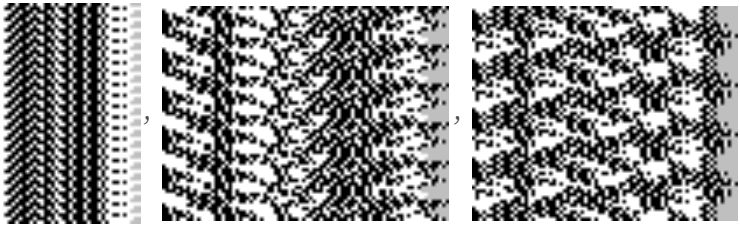

## Exploring Further

I don't know how far Emil Post got in exploring his tag system by hand a century ago. And I rather suspect that we've already gone a lot further here than he ever did. But what we've seen has just deepened the mystery of what tag systems can do. So far, every initial string we've tried has evolved to a cycle (or just terminated). But will this always happen? And how long can it take?

So far, the longest transient we've seen is 2141 steps—from the length-6 compressed string 111010:0. Length-7 and length-8 strings at most just "follow the same highway" in the state transition graph, and don't give longer transients. But at length 9 something different happens: 111111010:0 takes 24,552 steps to evolve a 6-cycle (with string length 12), with the lengths of intermediate (compressed) strings being:

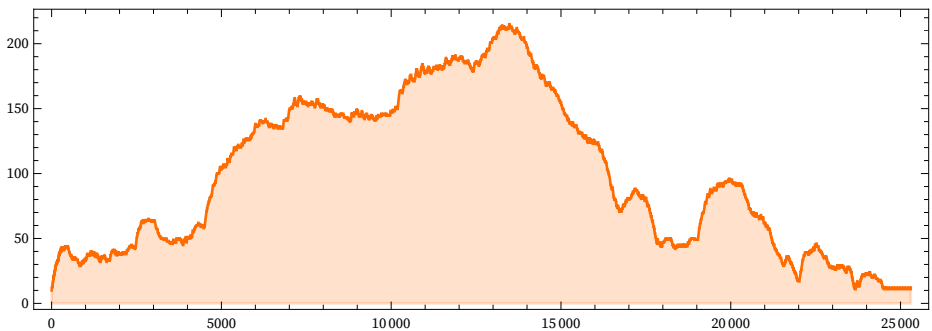

Plotting (from left to right) the actual elements in compressed strings in each "generation" this shows in more detail what's "going on inside":

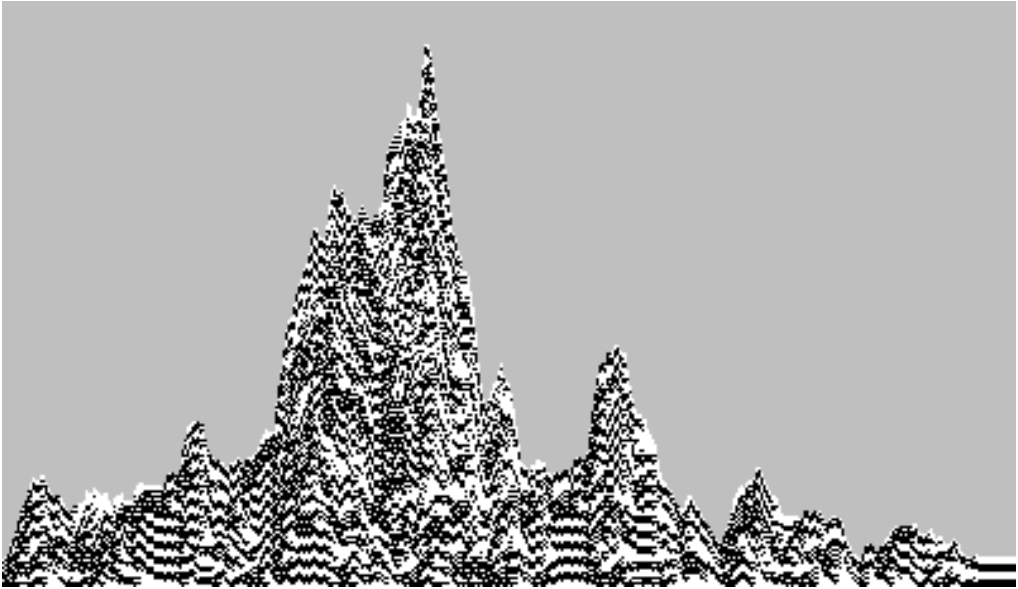

In systematically exploring what can happen in tag systems, it's convenient to specify initial compressed strings by converting their sequences of 1s and 0s to decimal numbers—but because our strings can have leading 0s we have to include the length, say as a prefix. So with this setup our length-9 "halting time winner" 111111010:0 becomes 9:506:0.

The next "winner" is 12:3962:0, which takes 253,456 steps to evolve to a 6-cycle:

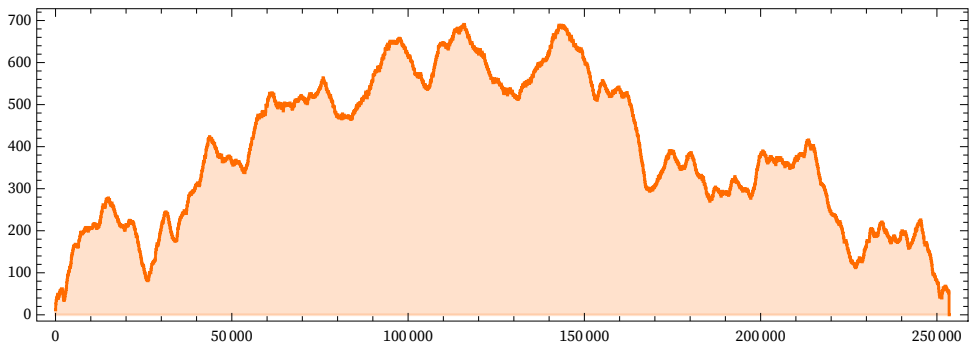

In generational form the explicit evolution in this case is:

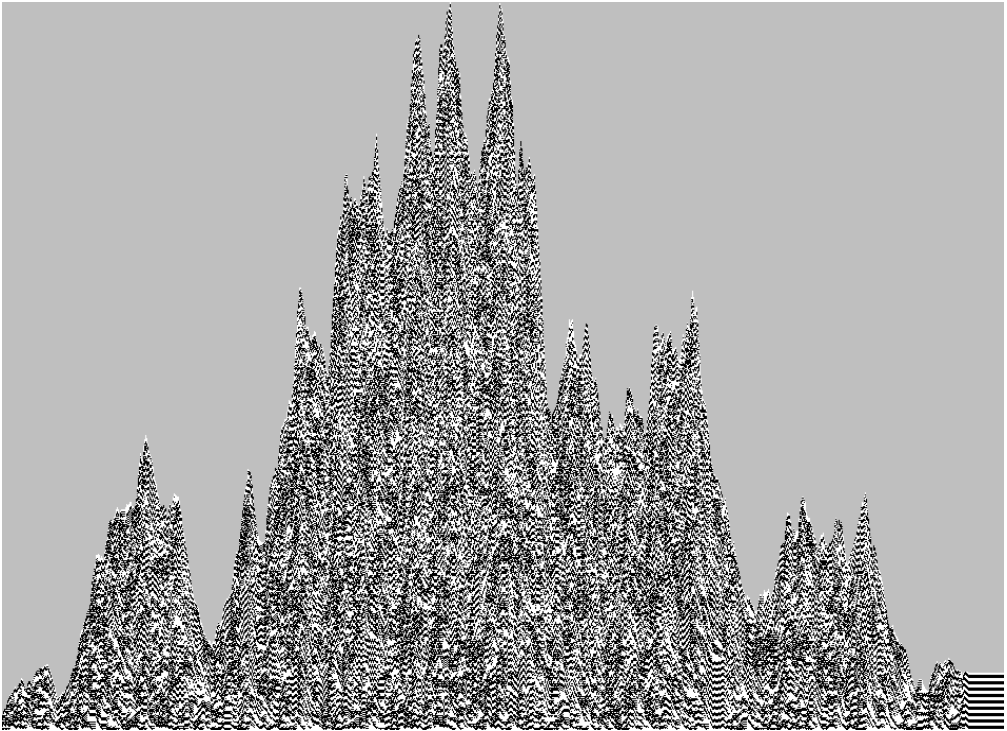

The first case to take over a million steps is 15:30166:0—which terminates after 20,858,103 steps:

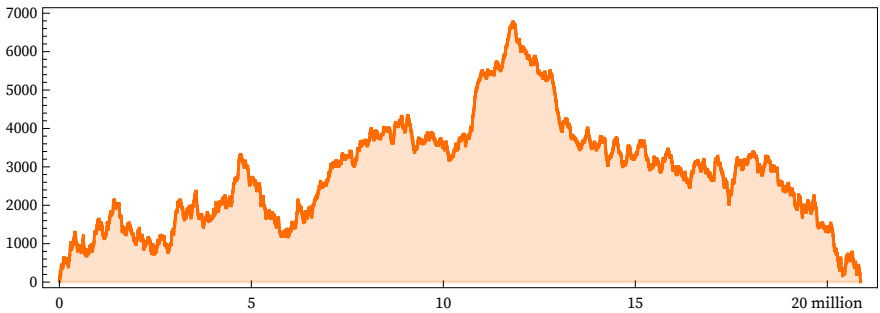

The first case to take over a billion steps is 20:718458:0—which leads to a 6-cycle after 2,586,944,112 steps:

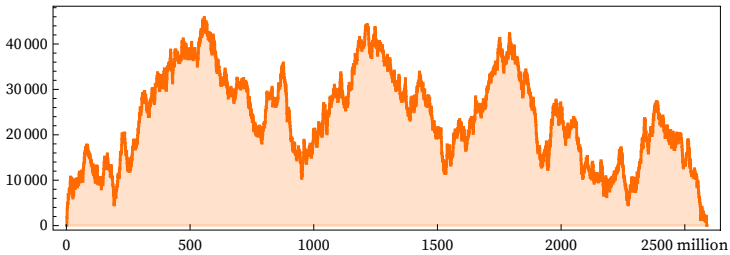

Here's table of all the "longest-so-far" winners through compressed initial length-28 strings (i.e. covering all $\approx 2 \times 10^{25}$ ordinary initial strings up to length 84):

| initial state | steps | cycle length |
|---|---:|---:|
| 4:14$_0$ | 419 | 0 |
| 6:58$_0$ | 2141 | 28 |
| 9:506$_0$ | 24 552 | 6 |
| 12:3962$_0$ | 253 456 | 6 |
| 13:5854$_0$ | 341 992 | 6 |
| 15:16 346$_0$ | 20 858 069 | 0 |
| 15:30 074$_0$ | 357 007 576 | 6 |
| 20:703 870$_0$ | 2 586 944 104 | 6 |
| 22:3 929 706$_0$ | 2 910 925 472 | 6 |
| 24:12 410 874$_0$ | 50 048 859 310 | 0 |
| 25:33 217 774$_0$ | 202 880 696 061 | 6 |
| 27:125 823 210$_0$ | 259 447 574 536 | 6 |
| 28:264 107 671$_2$ | 643 158 954 877 | 10 |

And here are their "size traces":

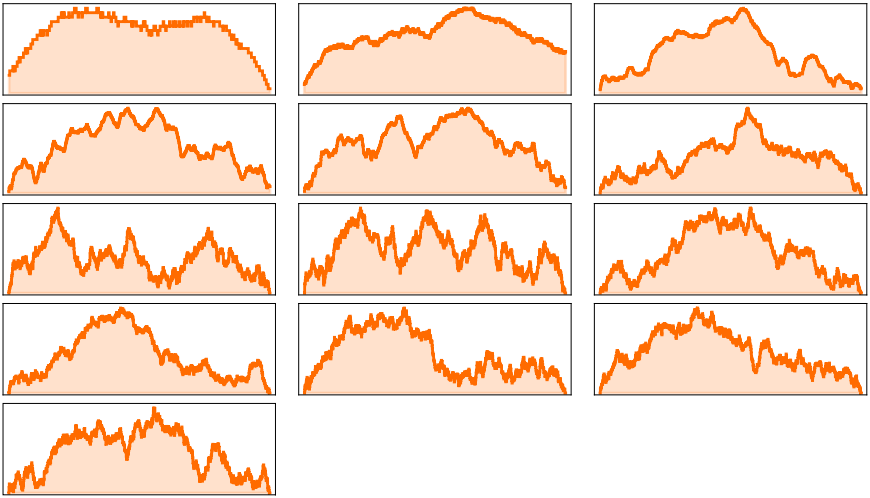

One notable thing here—that we'll come back to—is that after the first few cases, it's very difficult to tell the overall scale of these pictures. On the first row, the longest x axis is about 20,000 steps; on the last row it is about 600 billion.

But probably the most remarkable thing is that we now know that for all (uncompressed) initial strings up to length 75, the system always eventually evolves to a cycle (or terminates).

## Are They Like Random Walks?

Could the sequences of lengths in our tag system be like random walks? Obviously they can't strictly be random walks because given an initial string, each entire "walk" is completely determined, and nothing probabilistic or random is introduced.

But what if we look at a large collection of initial conditions? Could the ensemble of observed walks somehow statistically be like random walks? From the basic construction of the tag system we know that at each step the (uncompressed) string either increases or decreases in length by one element depending on whether its first element is 1 or 0.

But if we just picked increase or decrease at random here are two typical examples of ordinary random walks we'd get:

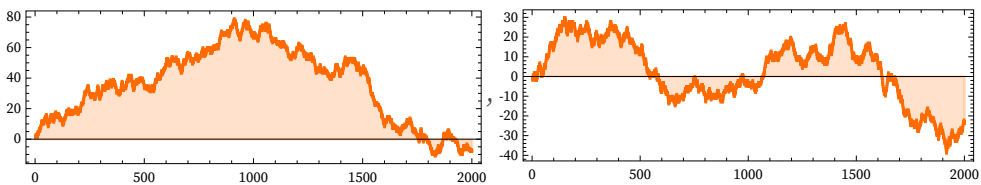

One very obvious difference from our tag system case is these walks can go below 0, whereas in the tag system case once one's reached something at least close to 0 (corresponding to a cycle), the walk stops. (In a market analogy, the time series ends if there's "bankruptcy" where the price hits 0.)

An important fact about random walks (at least in one dimension) is that with probability 1 they always eventually reach any particular value, like 0. So if our tag system behaved enough like a random walk, we might have an argument that it must "terminate with probability 1" (whatever that might mean given its discrete set of possible initial conditions).

But how similar can the sequence generated by a tag system actually be to an ordinary random walk? An important fact is that—beyond its initial condition—any tag system sequence must always consist purely of concatenations of the blocks 00 and 1101, or in other words, the sequence must be defined by a path through the finite automaton:

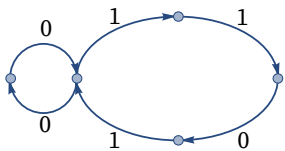

And from this we can see that—while all 2-grams and 3-grams can occur—the 4-grams 1111, 1100, 0101 and 0010 can never occur. In addition, if we assume that 0s and 1s occur with equal probability at the beginning of the string, then the blocks 00 and 1101 occur with equal probability, but the 3-grams 000, 011 occur with double the probabilities of the others.

In general the numbers of possible m-grams for successive m are 2, 4, 8, 12, 15, 20, 25, 33, 41, ... or for all m≥3:

$$\sum_{i}^{m} \text{Fibonacci}[\text{Ceiling}[\frac{i}{2}+2]] + 5 = \text{If}[\text{EvenQ}[m], 2\, \text{Fibonacci}[m/2 + 4], \text{Fibonacci}[(m + 11)/2]] - 1$$

Asymptotically this is $\approx \phi^{m/2}$ ---implying a limiting set entropy of $\log_2(\phi)/2 \approx 0.347$ per element. The relative frequencies of m-grams that appear (other than 0000...) are always of the form $2^a$. The following lists for each m the number of m-grams that appear at given multiplicities (as obtained from Flatten[DeBruijnSequence[{{0,0},{1,1,0,1}},m]]):

| m | 1 | 2 | 3 | 4 | 5 | 6 | 7 | 8 | 9 | 10 | 11 | 12 | 13 | 14 | 15 | 16 | 17 | 18 | 19 |
|---|---|---|---|---|---|---|---|---|---|---|---|---|---|---|---|---|---|---|---|
| multiplicity 3 | 0 | 0 | 0 | 1 | 0 | 1 | 0 | 1 | 0 | 1 | 0 | 1 | 0 | 1 | 0 | 1 | 0 | 1 | 0 |
| multiplicity 1 | 2 | 4 | 4 | 3 | 6 | 3 | 6 | 3 | 6 | 3 | 6 | 3 | 6 | 3 | 6 | 3 | 6 | 3 | 6 |
| multiplicity 2 | 0 | 0 | 4 | 7 | 9 | 11 | 17 | 15 | 25 | 19 | 33 | 23 | 41 | 27 | 49 | 31 | 57 | 35 | 65 |
| multiplicity 4 | 0 | 0 | 0 | 1 | 0 | 5 | 2 | 13 | 10 | 25 | 26 | 41 | 50 | 61 | 82 | 85 | 122 | 113 | 170 |
| multiplicity 8 | 0 | 0 | 0 | 0 | 0 | 0 | 0 | 1 | 0 | 6 | 2 | 19 | 12 | 44 | 38 | 85 | 88 | 146 | 170 |
| multiplicity 16 | 0 | 0 | 0 | 0 | 0 | 0 | 0 | 0 | 0 | 0 | 0 | 1 | 0 | 7 | 2 | 26 | 14 | 70 | 52 |
| multiplicity 32 | 0 | 0 | 0 | 0 | 0 | 0 | 0 | 0 | 0 | 0 | 0 | 0 | 0 | 0 | 0 | 1 | 0 | 8 | 2 |
| all m-grams | 2 | 4 | 8 | 12 | 15 | 20 | 25 | 33 | 41 | 54 | 67 | 88 | 109 | 143 | 177 | 232 | 287 | 376 | 465 |
| possible blocks | 2 | 4 | 8 | 16 | 32 | 64 | 128 | 256 | 512 | 1024 | 2048 | 4096 | 8192 | 16384 | 32768 | 65536 | 131072 | 262144 | 524288 |

(This implies a "p log p" measure entropy of below 0.1.)

So what happens in actual tag system sequences? Once clear of the initial conditions, they seem to quite accurately follow these probabilistic ("mean-field theory") estimates, though with various fluctuations. In general, the results are quite different from a pure ordinary random walk with every element independent, but in agreement with the estimates for a "00, 1101 random walk".

Another difference from an ordinary random walk is that our walks end whenever they reach a cycle—and we saw above that there are an infinite number of cycles, of progressively greater sizes. But the density of such "trap" states is small: among all size-n strings, only perhaps $1.5^{-n}$ of them lie on cycles.

The standard theory of random walks says, however, that in the limit of infinitely large strings and long walks, if there is indeed a random process underneath, these things will not matter: we'll have something that is in the same universality class as the ordinary ±1 random walk, with the same large-scale statistical properties.

But what about our tag systems that survive billions of steps before hitting 0? Could genuine random walks plausibly survive that long? The standard theory of first passage times (or "stopping times") tells us that the probability for a random walk starting at 0 to first reach x (or, equivalently, for a walk starting at x to reach 0) at time t is:

$$P(t) = \frac{x \exp\left(-\frac{x^2}{2t}\right)}{\sqrt{2\pi t^3}}$$

This shows the probability of starting from x and first reaching 0 as a function of the number of steps:

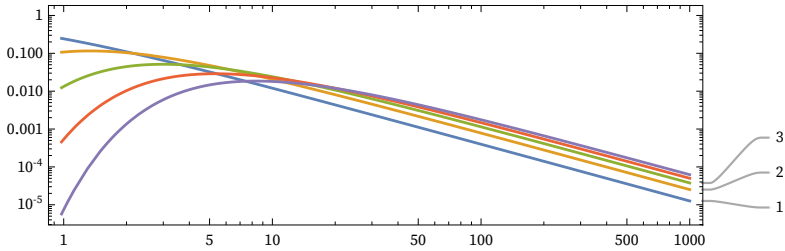

The most likely stopping time is $x^{2/3}$, but there is a long tail, and the probability of surviving for a time longer than t is:

$$\mathrm{erf}\left(\frac{x}{\sqrt{2t}}\right) \approx \sqrt{\frac{2}{\pi t}}\, x$$

How does this potentially apply to our systems? Assume we start from a string of (compressed) length n. This implies that the probability to survive for t steps (before "reaching x = 0") is about `Sqrt`[2/(`Pi` t)] n. But there are $3\, 2^n$ possible strings of length n. So we can roughly estimate that one of them might survive for about $18\, n^2\, 4^n / \mathrm{Pi}$ steps, or at least a number of steps that increases roughly exponentially with n.

And our results for "longest-so-far winners" above do in fact show roughly exponential increase with n (the dotted line is $4^{0.75\,n} \approx 2.8^n$):

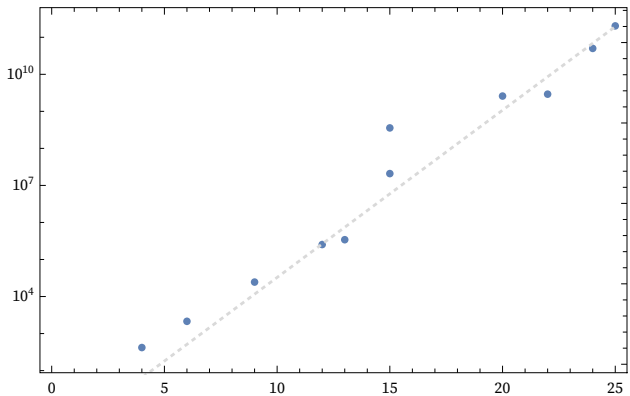

We can do a more detailed comparison with random walks by looking at the complete distribution of halting (AKA stopping) times for tag systems. Here are the results for all n = 15 and 25 initial strings:

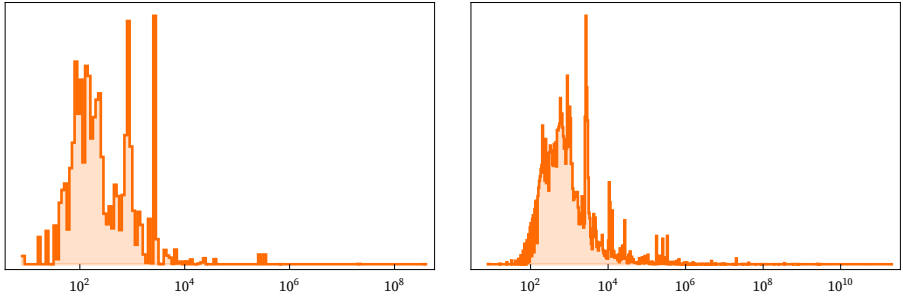

Plotting these on a log scale we get

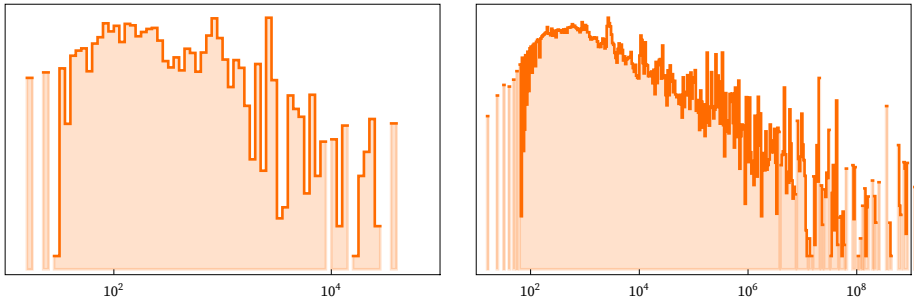

showing at least a rough approximation to the $1/\sqrt{t}$ behavior expected for a random walk.

In making distributions like these, we're putting together all the initial strings of length n, and asking about the statistical properties of this ensemble. But we can also imagine seeing whether initial strings with particular properties consistently behave differently from others. This shows the distribution of halting times as a function of the number of 1s in the initial string; no strong correlations are seen (here for n=20), even though at least at the beginning the presence of 1s leads to growth:

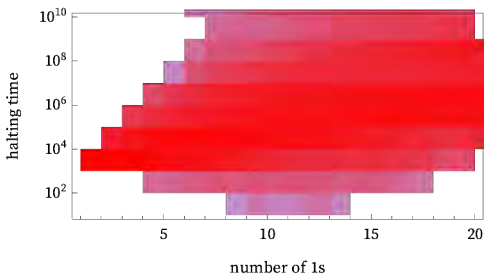

26 | Stephen Wolfram

## Analogies & Expectations

How should we think about what we're seeing? To me it in many ways just seems a typical manifestation of the ubiquitous phenomenon of computational irreducibility. Plenty of systems show what seems like random walk behavior. Even in rule 30, for example, the dividing line between regularity and randomness appears to follow a (biased) random walk:

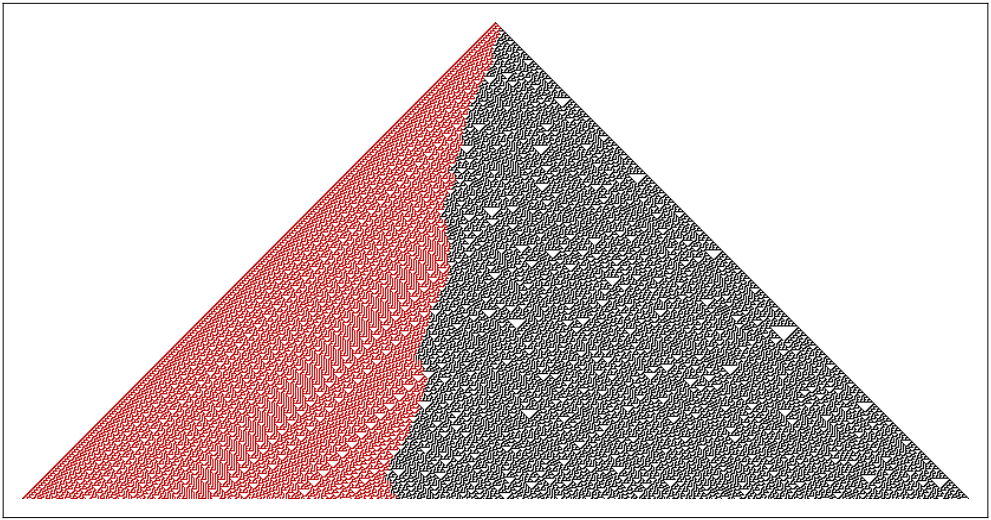

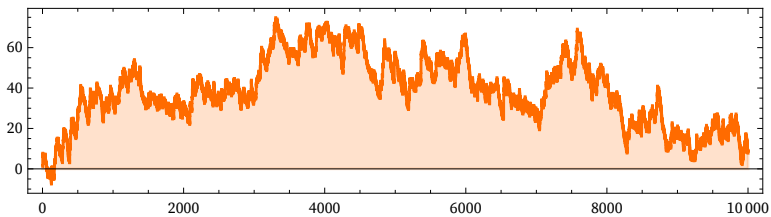

If we changed the initial conditions, we'd get a different random walk. But in all cases, we can think of the evolution of rule 30 as intrinsically generating apparent randomness, "seeded" by its initial conditions.

Even more directly analogous to our tag system are cellular automata whose boundaries show apparent randomness. An example is the k = 2, r = 3/2 rule 7076:

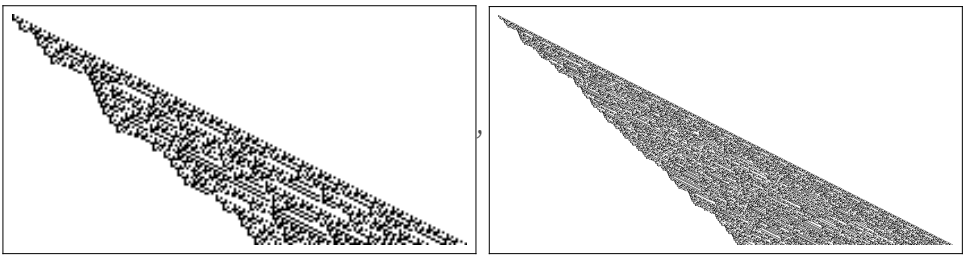

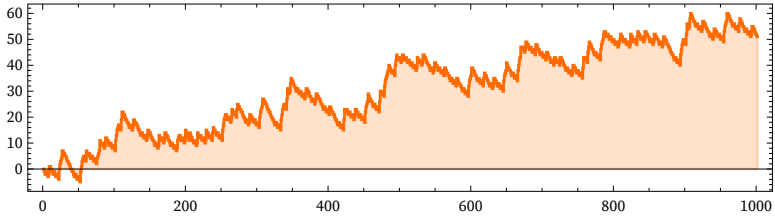

Will this pattern go on growing forever, or will it eventually become very narrow, and either enter a cycle or terminate entirely? This is analogous to asking whether our tag system will halt.

There are other cellular automata that show even more obvious examples of these kinds of questions. Consider the k = 3, r = 1 totalistic code 1329 cellular automaton. Here is its behavior for a sequence of simple initial conditions. In some cases the pattern dies out ("it halts"); in some cases it evolves to a (rather elaborate) period-78 cycle. And in one case here it evolves to a period-7 cycle:

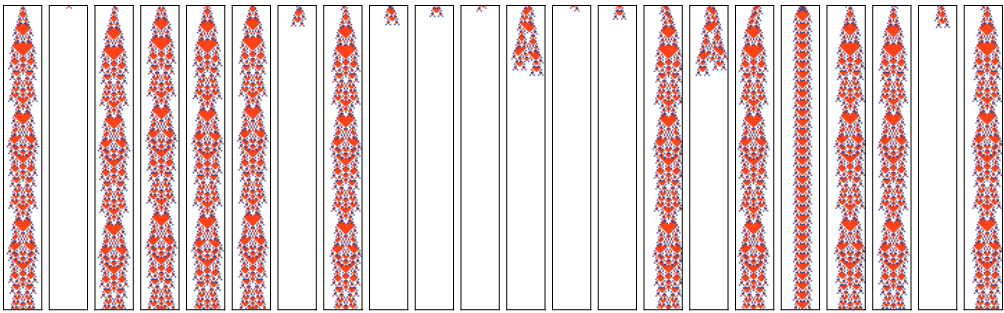

But is this basically all that can happen? No. Here are the various persistent structures that occur with the first 10,000 initial conditions—and we see that in addition to getting ordinary "cycles", we also get "shift cycles":

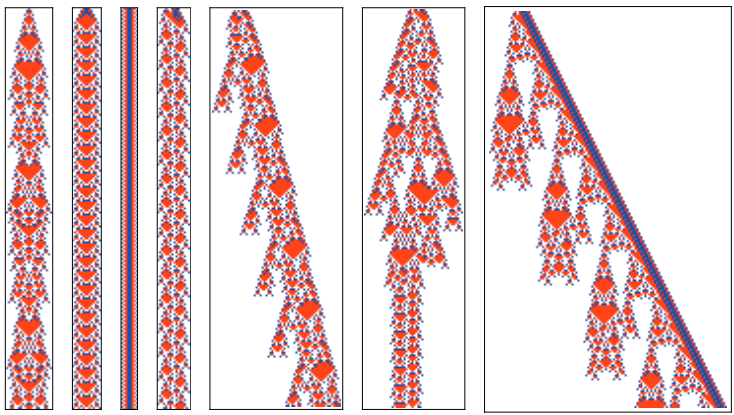

But if we go a little further, there's another surprise: initial condition 54,889 leads to a structure that just keeps growing forever—while initial condition 97,439 also does this, but in a much more trivial way:

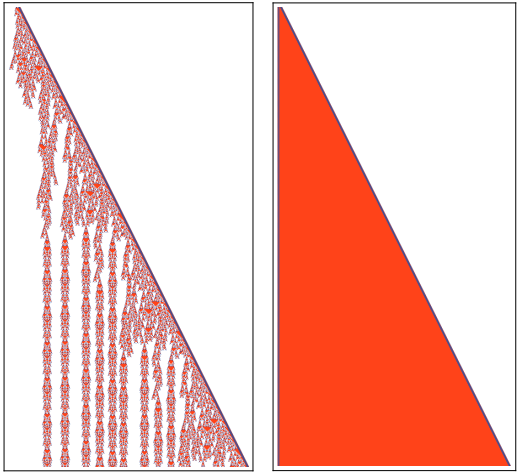

In our tag system, the analog of these might be particular strings that produce patterns that "obviously grow forever".

One might think that there could be a fundamental difference between a cellular automaton and a tag system. In a cellular automaton the rules operate in parallel, in effect connecting a whole grid of neighboring cells, while in a tag system the rules only specifically operate on the very beginning and end of each string.

But to see a closer analogy we can consider every update in the tag system as an "event", then draw a causal graph that shows the relationships between these events. Here is a simple case:

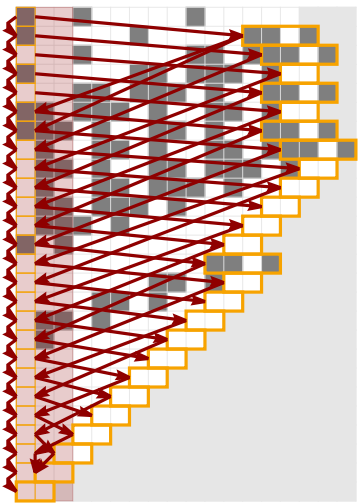

Extracting the pure causal graph we get:

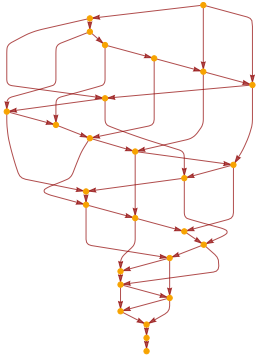

For the string 4:14:0 which takes 419 steps to terminate, the causal graph is:

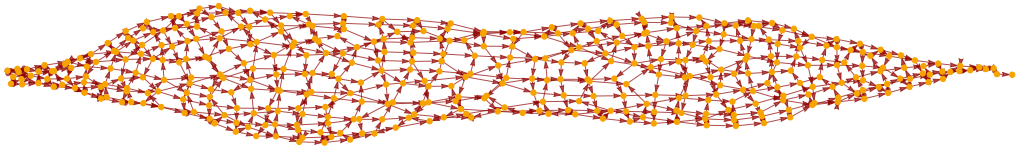

Or laid out differently, and marking expansion (1→1101) and contraction (0→00) events with red and blue:

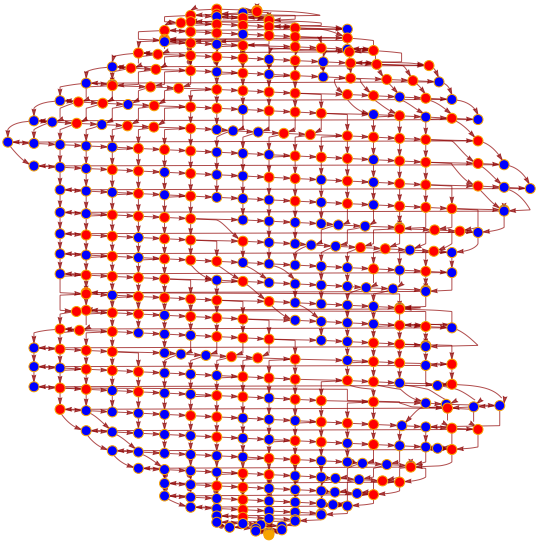

Here is the causal graph for the 2141-step evolution of 6:58:0

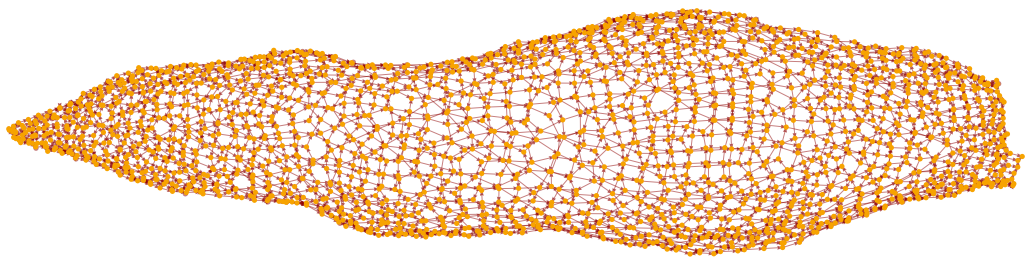

and what is notable is that despite the "spatial localization" of the underlying operation of the tag system, the causal graph in effect connects events in something closer to a uniform mesh.

## Connecting to Number Theory

When Emil Post was first studying tag systems a hundred years ago he saw them as the last hurdle in finding a systematic way to "solve all of mathematics", and in particular to solve all problems in number theory. Of course, they turned out to be a very big hurdle. But having now seen how complex tag systems can be, it's interesting to go back and connect again with number theory.

It's straightforward to convert a tag system into something more obviously number theoretical. For example, if one represents each string of length n by a pair of integers {n,i} in which the binary digits of i give the elements of the string, then each step in the evolution can be obtained from:

TagStep[{n_, i_}] :=
  With[{j = $2^{n-1}$ FractionalPart[$\frac{8\,i}{2^n}$]}, If[i < $2^{n-1}$, {n – 1, j}, {n + 1, 4 j + 13}]]

Starting from the 4:14:0 initial condition (here represented in uncompressed form by {12, 2336}) the first few steps are then:

NestList[TagStep, {12, 2336}, 10]

{{12, 2336}, {13, 4621}, {14, 8413}, {15, 3549}, {14, 14 196},
  {15, 30 541}, {16, 29 917}, {15, 21 364}, {16, 14 157}, {15, 23 860}, {16, 54 093}}

For compressed strings, the corresponding form is:

TagStep[{n_, i_, p_}] := With[{j = $2^n$ FractionalPart[$\frac{i}{2^{n-1}}$]},
  If[i < $2^{n-1}$, {{n, j, 2}, {n – 1, $\frac{j}{2}$, 0}, {n, j, 1}}, {{n + 1, 2 j + 3, 1}, {n, j, 2}, {n, j + 1, 0}}][[p + 1]]]

There are different number theoretical formulations one can imagine, but a core feature is that at each step the tag system is making a choice between two arithmetic forms, based on some essentially arithmetic property of the number obtained so far. (Note that the type of condition we have given here can be further "compiled" into "pure arithmetic" by extracting it as a solution to a Diophantine equation.)

A widely studied system similar to this is the Collatz or 3n + 1 problem, which generates successive integers by applying the function:

$n \mapsto \text{If}[\text{EvenQ}[n], n/2, 3n+1]$

Starting, say, from 27, the sequence of numbers obtained is 27, 82, 41, 124, 62, 31, ...

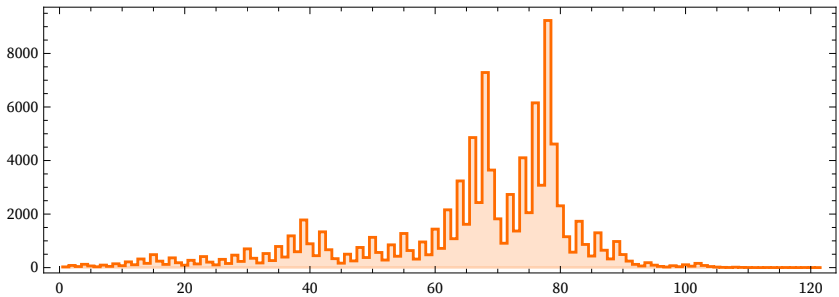

where after 110 steps the system reaches the cycle 4, 2, 1, 4, 2, 1, .... As a closer analog to the plots for tag systems that we made above we can instead plot the lengths of the successive integers, represented in base 2:

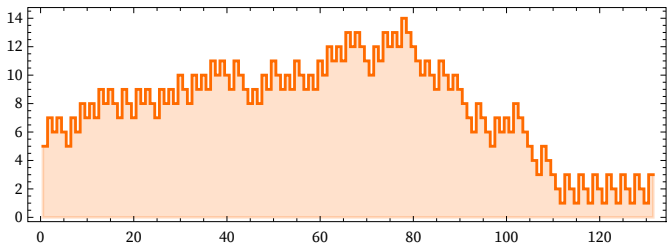

The state transition graph starting from integers up to 10 is

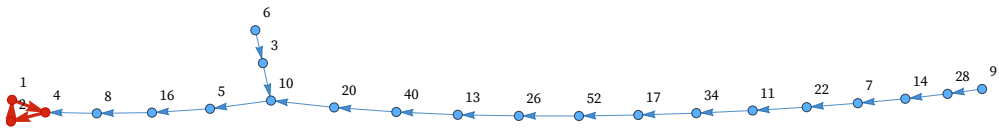

and up to 1000 it is:

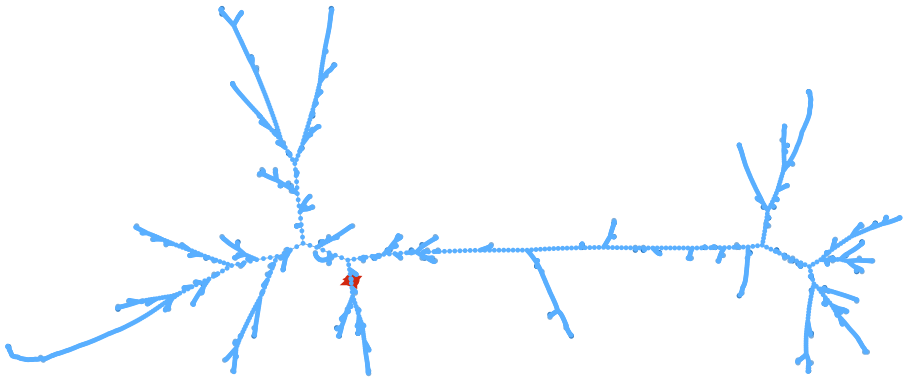

Unlike for Post's tag system, there is only one connected component (and one final cycle), and the "highways" are much shorter. For example, among the first billion initial conditions, the longest transient is just 986 steps. It occurs for the initial integer 670617279—which yields the following sequence of integer lengths:

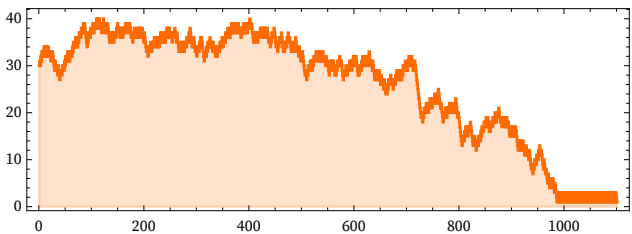

Despite a fair amount of investigation since the 1930s, it's still not known whether the 3n + 1 problem always terminates on its standard cycle—though this is known to be the case for all integers up to $10^{16}$.

For Post's tag system the most obvious probabilistic estimate suggests that the sequence of string lengths should follow an unbiased random walk. For the 3n + 1 problem, a similar analysis suggests a random walk with an average bias of $\log_2((3/4)^{1/3}) \approx$ -0.14 binary digits per step, as suggested by this collection of walks from initial conditions $10^8 + k$:

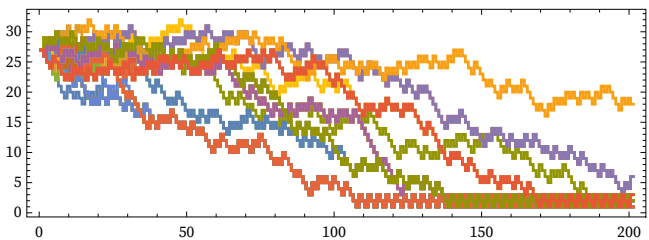

The rule (discussed in A New Kind of Science)

n ⟼ If[EvenQ[n], n/2, 5 n + 1]

instead implies a bias of +0.11 digits per step, and indeed most initial conditions lead to growth:

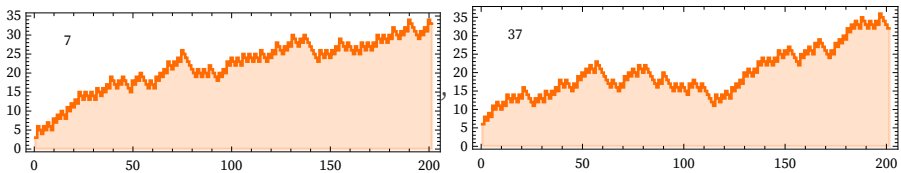

But there are still some that—even though they grow for a while—have "fluctuations" that cause them to "crash" and end up in cycles:

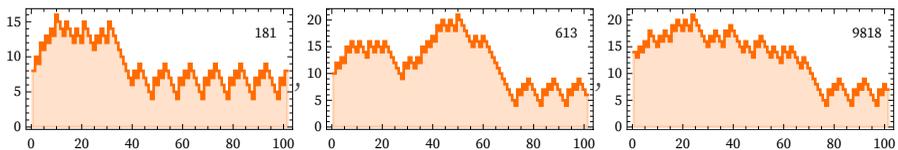

What is the "most unbiased" a n + b system? If we consider mod 3 instead of mod 2, we have systems like:

$$n \mapsto \{n, a_1 n + b_1, a_2 n + b_2\}_{[[Mod[n,3]+1]]} / 3$$

We need $a_i\, n + b_i$ to be divisible by 3 when $n = i$ mod 3. In our approximation, the bias will be $\log_2(a_1 a_2 / 27)$. This is closest to zero (with value +0.05) when $a_i$ are 4 and 7. An example of a possible iteration is then:

$$n \mapsto \{n, 4 n + 2, 7 n + 1\}_{[[Mod[n,3]+1]]} / 3$$

Starting from a sequence of initial conditions this clearly shows less bias than the 3n + 1 case:

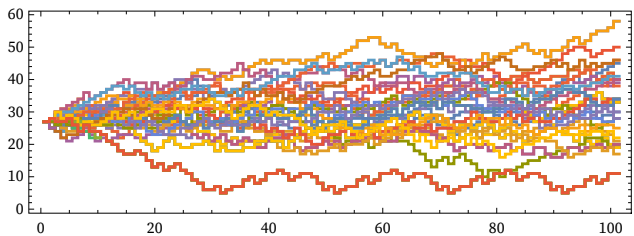

Here are the halting times for initial conditions up to 1000:

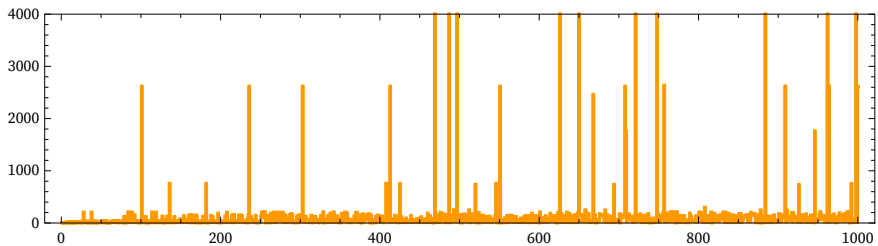

Most initial conditions quickly evolve to cycles of length 5 or 20. But initial condition 101 takes 2604 steps to reach the 20-cycle:

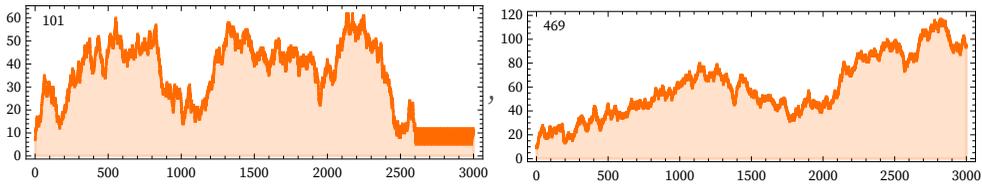

And initial condition 469 does not appear to reach a cycle at all—and instead appears to systematically grow at about 0.018 bits per step:

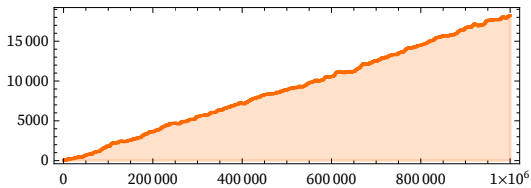

In other words, unlike the 3n + 1 problem—or our tag system—this iteration usually leads to a cycle, but just sometimes appears to "escape" and continue to increase, presumably forever.

(In general, for modulus m, the minimum bias will typically be $\log_2((m^m \pm 1)/m^m)$, and the "smoothest" iterations will be ones whose multipliers involve similar-sized factors of numbers close to $m^m$. For m = 4, for example, {n, 3n – 3, 5n – 2, 17n + 1} is the best.)

One might wonder how similar our tag system—or the 3n + 1 problem—is to classic unsolved problems in number theory, like the Riemann Hypothesis. In essence the Riemann Hypothesis is an assertion about the statistical randomness of primes, normally stated in terms of complex zeroes of the Riemann zeta function, or equivalently, that all the maxima of RiemannSiegelZ[t] (for any value of t) lie above the axis:

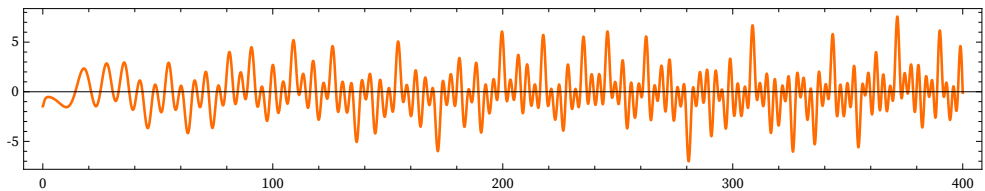

But it's known (thanks to extensive work by Yuri Matiyasevich) that an equivalent—much more obviously integer-related—statement is that

$$\frac{(2n+3)!!}{15} - (2n-2)!!\, \text{PrimePi}[n]^2\, ((\text{BitLength}[\text{Fold}[\text{LCM}, \text{Range}[n]]] - 1) \sum_{k=1}^{n-1} (-1)^{k+1} k^{-1} - n)$$

is positive for all positive n. And this then turns out to be equivalent to the surprisingly simple statement that the iteration

NestWhile[x ⟼ {2 $x_{[\![2]\!]}$ $x_{[\![1]\!]}$ – 4 (–1)$^{x_{[\![2]\!]}}$ $x_{[\![5]\!]}$, $x_{[\![2]\!]}$ + 1, ($x_{[\![2]\!]}$ + 1) $x_{[\![3]\!]}$/GCD[$x_{[\![2]\!]}$ + 1, $x_{[\![3]\!]}$],
    If[GCD[$x_{[\![2]\!]}$ + 1, $x_{[\![3]\!]}$] == 1, $x_{[\![4]\!]}$ + 1, $x_{[\![4]\!]}$], $x_{[\![6]\!]}$, (2 $x_{[\![2]\!]}$ + 2) $x_{[\![6]\!]}$, (2 $x_{[\![2]\!]}$ + 5) $x_{[\![7]\!]}$},
  {1, 1, 1, 0, 0, 1, 1}, x ⟼ $x_{[\![7]\!]}$ > $x_{[\![4]\!]}$$^2$ ($x_{[\![1]\!]}$ (BitLength[$x_{[\![3]\!]}$] – 1) – $x_{[\![6]\!]}$)]

will never terminate.

For successive n the quantity above is given by:

{1, 9, 127, 981, 16 209, 207 135, 3 821 007, 55 666 269, 1 250 264 481, 26 427 261 303}

At least at the beginning the numbers are definitely positive, as the Riemann Hypothesis would suggest. But if we ask about the long-term behavior we can see something of the complexity involved by looking at the differences in successive ratios:

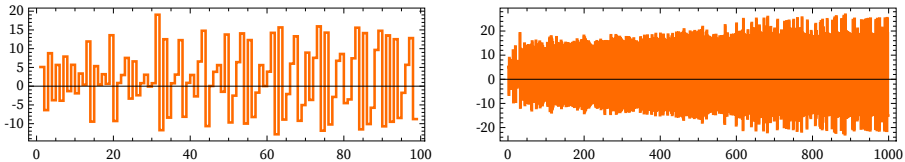

The Riemann Hypothesis effectively says that there aren't too many negative differences here.

## Other Tag Systems

So far we've been talking specifically about Emil Post's particular 00, 1101 tag system. But as Post himself observed, one can define plenty of other tag systems—including ones that involve not just 0 and 1 but any number of possible elements (Post called the number of possible elements $\mu$, but I'll call it k), and delete not just 3 but any number of elements at each step (Post called this $v$, but I'll call it r).

It's easy to see that rules which delete only one element at each step (r = 1) cannot involve real "communication" (or causal connections) between different parts of the string, and must be equivalent to neighbor-independent substitution systems—so that they either have trivial behavior, or grow without bound to produce at most highly regular nested sequences. (0→01, 1→10 will generate the Thue–Morse string, while 0→01, 1→0 will generate the Fibonacci string.)

Things immediately get more complicated when two elements are deleted at each step (r = 2). Post correctly observed that with just 0 and 1 (k = 2) there are no rules that show the kind of sometimes-expanding, sometimes-contracting behavior of his 00, 1101 rule. But back in 2007—as part of a live experiment at our annual Summer School—I looked at the r = 2 rule 0→1, 1→110. Here's what it does starting with 10:

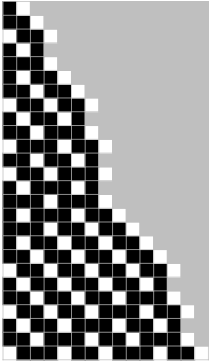

And here's how the sequence of string lengths behaves:

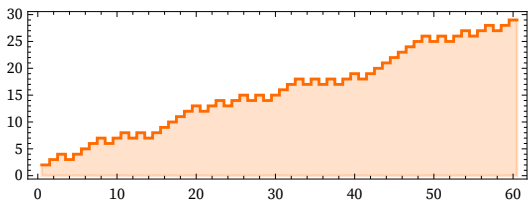

If we assume that 0 and 1 appear randomly with certain probabilities, then a simple calculation shows that 1 should occur about $1+\sqrt{2} \approx 2.41$ times as often as 0, and the string should grow an average of $\sqrt{2}-1 \approx 0.41$ elements at each step. So "detrending" by this, we get:

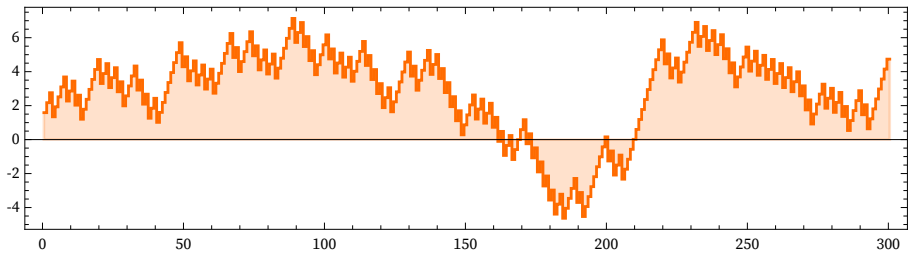

Continuing for more steps we see a close approximation to a random walk:

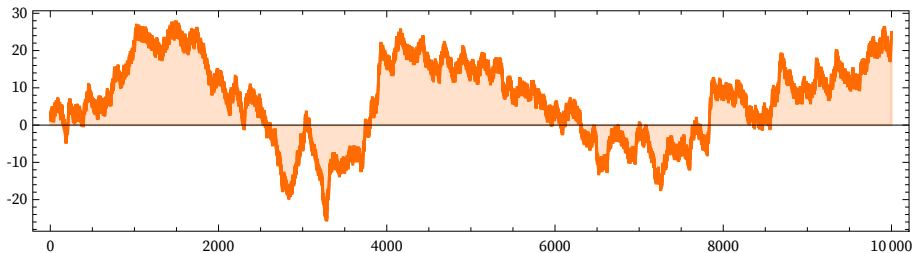

So just like with Post's 00, 1101 rule—and, of course, with rule 30 and all sorts of other systems in the computational universe—we have here a completely deterministic system that generates what seems like randomness. And indeed among tag systems of the type we're discussing here this appears to be the very simplest rule that shows this kind of behavior.

But does this rule show the same kind of growth from all initial conditions? It can show different random sequences, for example here for initial conditions 5:17 and 7:80:

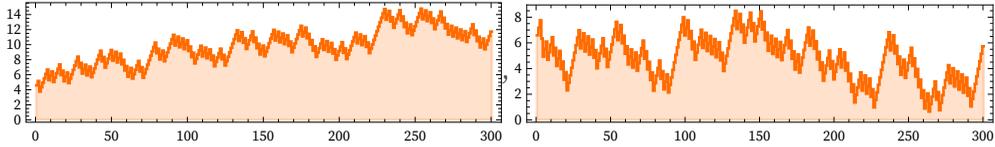

And sometimes it just immediately enters a cycle. But it has some "surprises" too. Like with initial condition 9:511 (i.e. 111111111) it grows not linearly, but like $\sqrt{t}$ (shown here without any detrending):

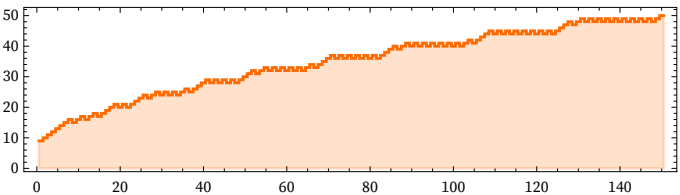

But what about a tag system that doesn't seem to "typically grow forever"? When I was working on A New Kind of Science I studied generalized tag systems that don't just look at their first elements, but instead use the whole block of elements they're deleting to determine what elements to add at the end (and so work in a somewhat more "cellular-automaton-style" way).

One particular rule that I showed in A New Kind of Science (as case (c) on page 94) is:

{00 → 0, 10 → 101, 01 → 000, 11 → 011}

Starting with 11 this rule gives

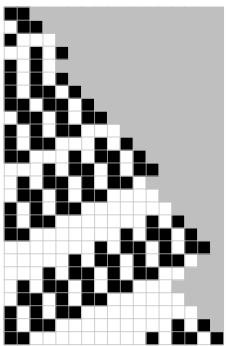

and grows for a while—but then terminates after 289 steps:

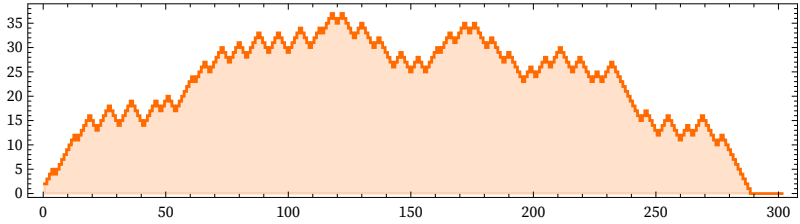

The corresponding generational evolution is:

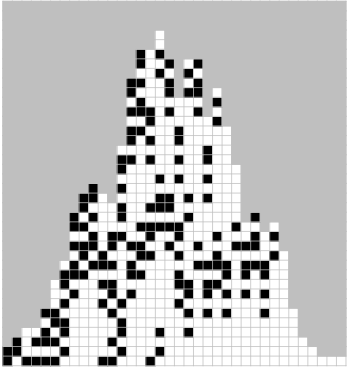

(Note that the kind of "phase decomposition" that we did for Post's tag system doesn't make sense for a block tag system like this.)

Here are the lengths of the transients+cycles for possible initial conditions up to size 7:

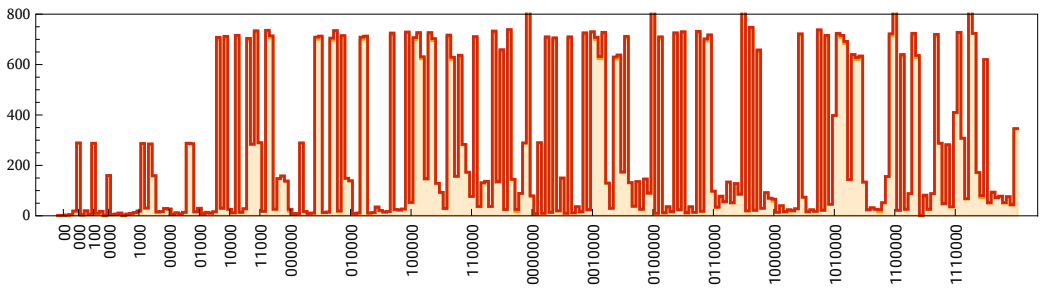

This looks more irregular—and "livelier"—than the corresponding plot for Post's tag system, but not fundamentally different. At size 5 the initial string 11010 (denoted 5:12) yields

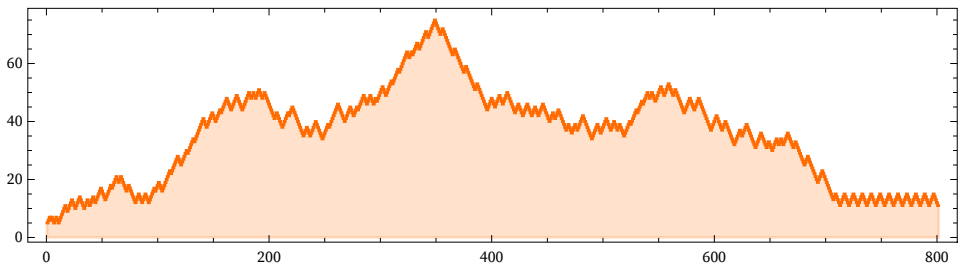

which terminates after 706 steps in a length-8 cycle. Going further one sees a sequence of progressively longer transients:

| initial state | steps | cycle length |
|---|---|---|
| 2:3 | 288 | 1 |
| 5:12 | 700 | 8 |
| 6:62 | 4184 | 1 |
| 8:175 | 20 183 | 8 |
| 9:345 | 26 766 | 1 |
| 9:484 | 51 680 | 8 |
| 10:716 | 100 285 | 1 |
| 10:879 | 13 697 828 | 8 |
| 13:7620 | 7 575 189 088 | 1 |
| 17:85 721 | 14 361 319 032 | 8 |

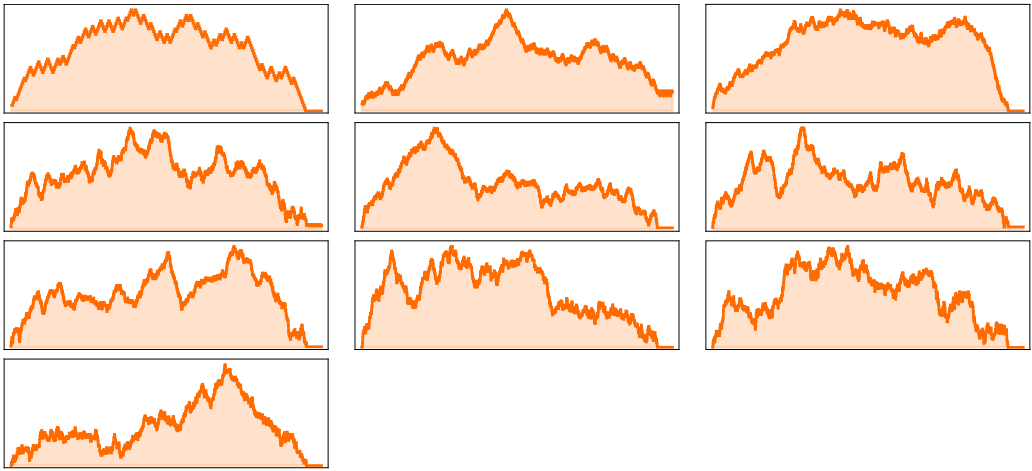

But like with Post's tag system, the system always eventually reaches a cycle (or terminates)—at least for all initial strings up to size 17. But what will happen for the longest initial strings is not clear, and the greater "liveliness" of this system relative to Post's suggests that if exotic behavior occurs, it will potentially do so for smaller initial strings than in Post's system.

Another way to generalize Post's 00, 1101 tag system is to consider not just elements 0, 1, but, say, 0, 1, 2 (i.e. k = 3). And in this case there is already complex behavior even with rules that consider just the first element, and delete two elements at each step (r = 2).

As an example, consider the rule:

{0 → 0, 1 → 02, 2 → 211}

Starting, say, with 101 this gives

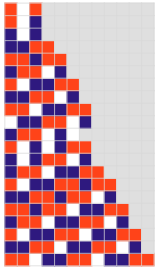

which terminates after 74 steps:

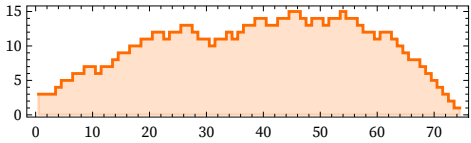

Here are the lengths of transients+cycles for this rule up to length-6 initial (ternary) strings:

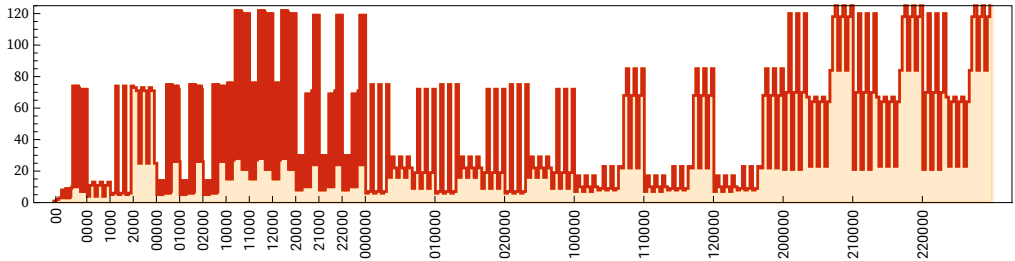

The initial string 202020 (denoted 6:546, where now this indicates ternary rather than binary) terminates after 6627 steps

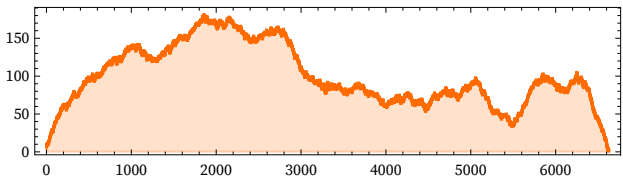

with (phase-reduced) generational evolution:

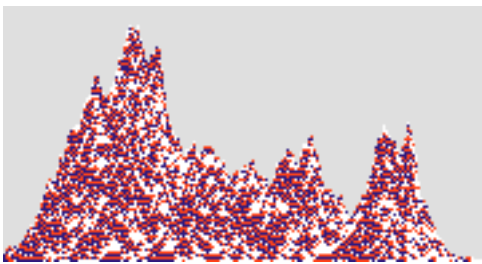

And once again, the overall features of the behavior are very similar to Post's system, with the longest halting times seen up to strings of length 14 being:

| initial state | steps | cycle length |
|---|---:|---:|
| $3{:}10_0$ | 74 | 0 |
| $5{:}91_0$ | 122 | 0 |
| $6{:}546_0$ | 6627 | 0 |
| $9{:}499_0$ | 9353 | 0 |
| $9{:}610_0$ | 12 789 | 0 |
| $9{:}713_0$ | 20 175 | 0 |
| $9{:}1214_0$ | 175 192 | 0 |
| $9{:}18\,787_0$ | 336 653 | 0 |
| $10{:}17\,861_0$ | 519 447 | 0 |
| $10{:}29\,524_0$ | 21 612 756 | 6 |
| $10{:}52\,294_0$ | 85 446 023 | 0 |
| $11{:}93\,756_0$ | 377 756 468 | 6 |
| $12{:}412\,474_0$ | 30 528 772 851 | 0 |

But what about other possible rules? As an example, we can look at all 90 possible k = 3, r = 2 rules of the form 0→_, 1→__, 2→___ in which the right-hand sides are "balanced" in the sense that in total they all contain two 0s, 1s and 2s. This shows the evolution (for 100 steps) for each of these rules that has the longest transient for any initial string with less than 7 elements:

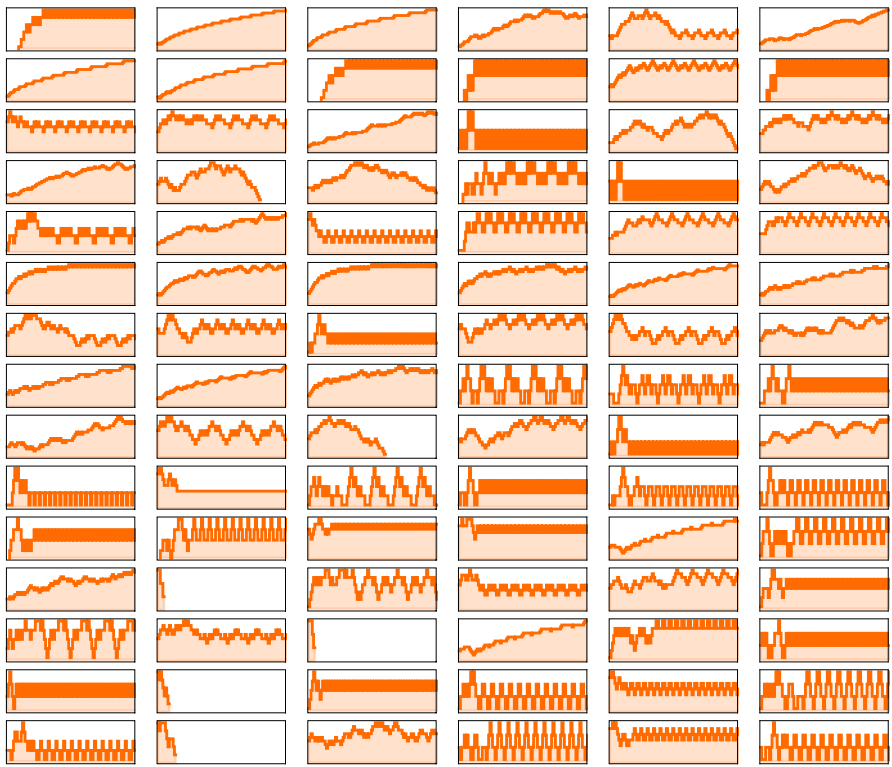

Many lead quickly to cycles or termination. Others after 100 steps seem to be growing irregularly, but all the specific evolutions shown here eventually halt. There are peculiar cases, like 0→0, 1→02, 2→112 which precisely repeats the initial string 20 after 18,255 steps:

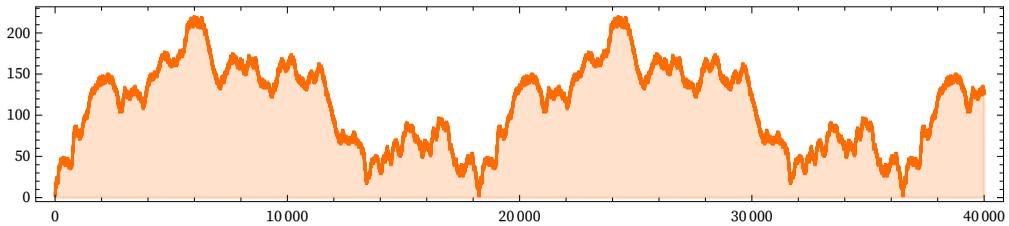

And then there are cases like 0→0, 1→01, 2→212, say starting from 200020, which either halt quickly, or generate strings of ever-increasing length (here like $\sqrt{t}$ ) and can easily be seen never to halt:

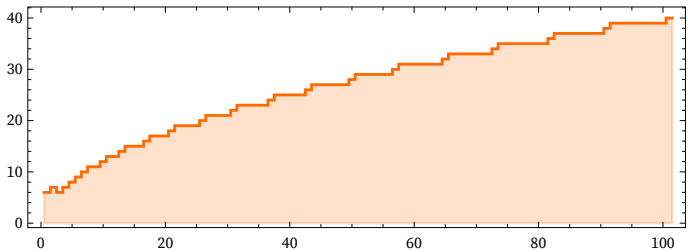

(By the way, the situation with "non-balanced" k = 3 rules is not fundamentally different from balanced ones; 0→0, 1→22, 2→102, for example, shows very "Post-like" behavior.)

The tag systems we've been discussing are pretty simple. But an even simpler version considered in A New Kind of Science are what I called cyclic tag systems. In a cyclic tag system one removes the first element of the string at each step. On successive steps, one cycles through a collection of possible blocks to add, adding one if the deleted element was a 1 (and otherwise adding nothing).

If the possible blocks to add are 111 and 0, then the behavior starting from the string 1 is as follows

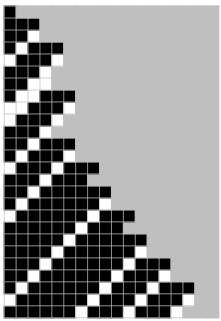

with the lengths "detrended by t/2" behaving once again like an approximate random walk:

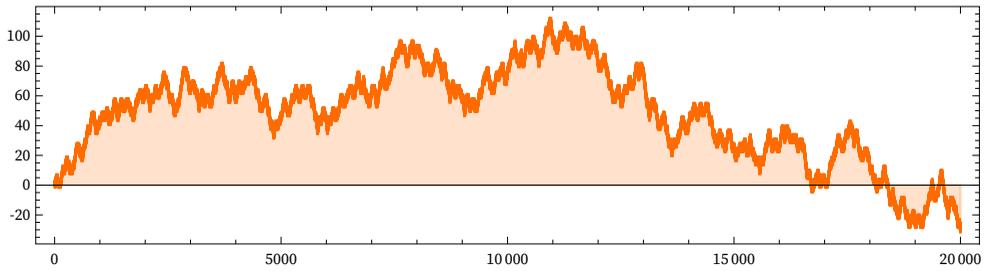

With cycles of just 2 blocks, one typically sees either quick cycling or termination, or what seems like obvious infinite growth. But if one allows a cycle of 3 blocks, more complicated halting behavior becomes possible.

Consider for example 01, 0, 011. Starting from 0111 one gets

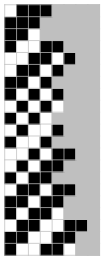

with the system halting after 169 steps:

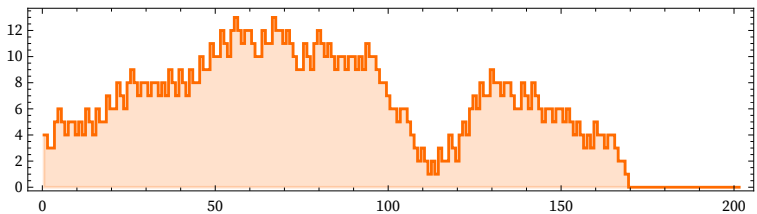

Here are the transient+cycle times for initial strings up to size 8 (the system usually just terminates, but for example 001111 goes into a cycle of length 18):

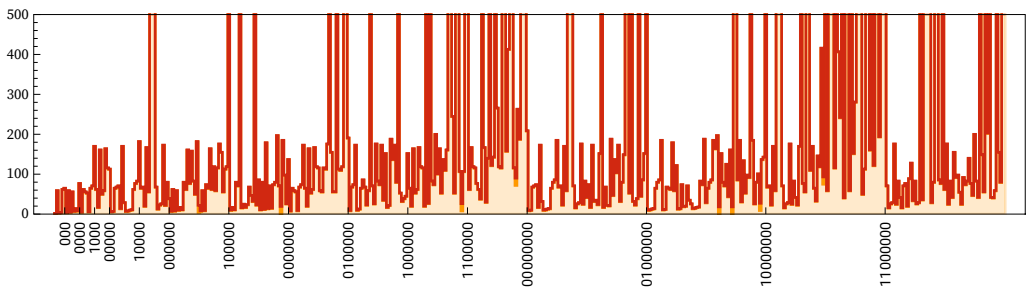

The behavior of the longest-to-halt-so-far "winners" are again similar to what we have seen before---except perhaps for the rather huge jump in halting time at length 13---that isn't surpassed until size 16:

| initial state | steps |
|---|---|
| 1:1 | 59 |
| 4:7 | 169 |
| 5:21 | 1259 |
| 7:126 | 6470 |
| 10:687 | 134 318 |
| 13:7655 | 10 805 957 330 |

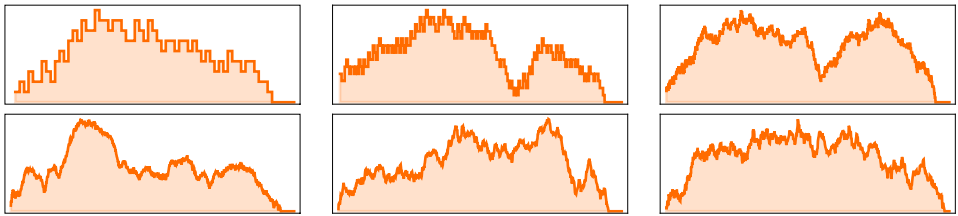

## What Can It Compute?

When Post originally invented tag systems in 1920 he intended them as a string-based idealization of the operations in mathematical proofs. But a decade and a half later, once Turing machines were known, it started to be clear that tag systems were better framed as being computational systems. And by the 1940s it was known that at least in principle string-rewriting systems of the kind Post used were capable of doing exactly the same types of computations as Turing machines—or, as we would say now, that they were computation universal.

At first what was proved was that a fairly general string-rewriting system was computation universal. But by the early 1960s it was known that a tag system that looks only at its first element is also universal. And in fact it's not too difficult to write a "compiler" that takes any Turing machine rule and converts it to a tag system rule—and page 670 of A New Kind of Science is devoted to showing a pictorial example of how this works:

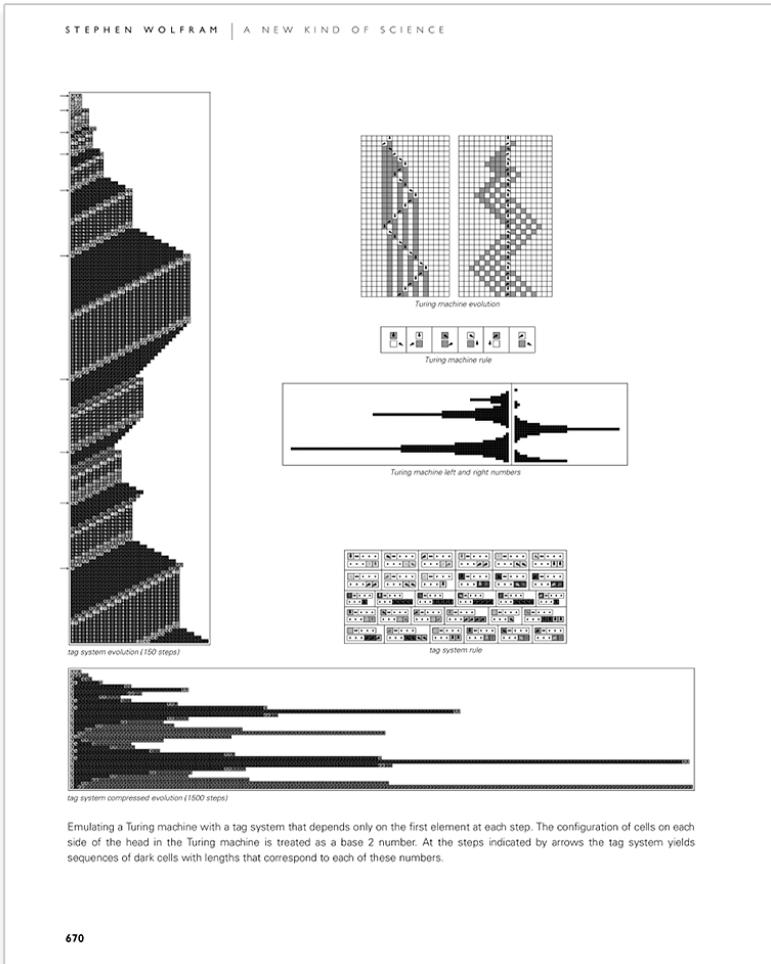

For example we can take the simplest universal Turing machine (which has 2 states and 3 colors) and compile it into a 2-element-deletion tag system with 32 possible elements (the ones above 9 represented by letters) and rules:

{0 → 32, 1 → 65, 2 → 11, 3 → 0, 4 →  , 5 → 00, 6 → 11, 7 → 00, 8 → ba, 9 → ed, a → 9999, b → 8,
  c → 8, d → 8, e → 9999, f → 8888, g → ji, h → ml, i → 99hh, j → ggoooo, k → ggoo, l → ggoooo,
  m → 9999h, n →  , o → rq, p → ut, q → p, r → oooo, s → oooo, t → oooo, u → p, v → o}

But what about a tag system like Post's 00, 1101 one—with much simpler rules? Could it also be universal?

Our practical experience with computers might make us think that to get universality we would necessarily have to have a system with complicated rules. But the surprising conclusion suggested by the Principle of Computational Equivalence is that this is not correct—and that instead essentially any system whose behavior is not obviously simple will actually be capable of universal computation.

For any particular system it's usually extremely difficult to prove this. But we now have several examples that seem to validate the Principle of Computational Equivalence—in particular the rule 110 cellular automaton and the 2,3 Turing machine. And this leads us to the conjecture that even tag systems with very simple rules (at least ones whose overall behavior is not obviously simple) should also be computation universal.

How can we get evidence for this? We might imagine that we could see a particular tag system "scanning over" a wide range of computations as we change its initial conditions. Of course, computation universality just says that it must be possible to construct an initial condition that performs any given computation. And it could be that to perform any decently sophisticated computation would require an immensely complex initial condition, that would never be "found naturally" by scanning over possible initial conditions.

But the Principle of Computational Equivalence actually goes further than just saying that all sorts of systems can in principle do sophisticated computations; it says that such computations should be quite ubiquitous among possible initial conditions. There may be some special initial conditions that lead to simple behavior. But other initial conditions should produce behavior that corresponds to a computation that is in a sense as sophisticated as any other computation.

And a consequence of this is that the behavior we see will typically be computationally irreducible: that in general there will be no way to compute its outcome much more efficiently than just by following each of its steps. Or, in other words, when we observe the system, we will have no way to computationally reduce it—and so its behavior will seem to us complex.

So when we find behavior in tag systems that seems to us complex—and that we do not appear able to analyze or predict—the expectation is that it must correspond to a sophisticated computation, and be a sign that the tag system follows the Principle of Computational Equivalence and is computation universal.

But what actual computations do particular tag systems do? Clearly they do the computations that are defined by their rules. But the question is whether we can somehow also interpret the overall computations they do in terms of familiar concepts, say in mathematics or computer science.

Consider for example the 2-element-deletion tag system with rules 1→111. Starting it off with 11 we get

```
11
111
1111
11111
111111
1111111
```

and we can see that the tag in effect just "counts up in unary". (The 1-element-deletion rule 1→11 does the same thing.)

Now consider the tag system with rules:

{1 → 22, 2 → 1111}

Starting it with 11 we get

11
22
1111
1122
2222
221111
11111111
11111122
11112222

or more pictorially (red is 1, blue is 2):

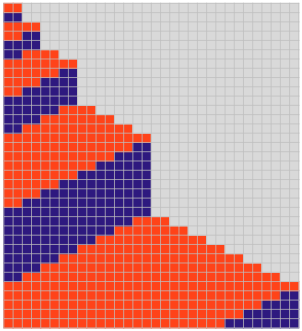

But now look at steps where strings of only 1s appear. The number of 1s in these strings forms the sequence

{2, 4, 8, 16, 32, 64, 128, 256, 512}

of successive powers of 2. (The 1-element-deletion rule 1→2, 2→11 gives the same sequence.)

The rule

{1 → 22, 2 → 111}

starting from 11 yields instead

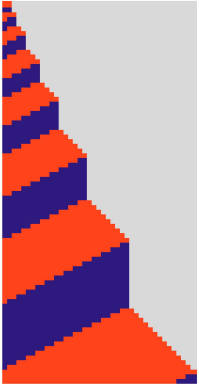

and now the lengths of the sequences of 1s form the sequence:

{2, 3, 5, 8, 12, 18, 27, 41, 62, 93, 140, 210, 315, 473, 710, 1065, 1598, 2397, 3596}

This sequence is not as familiar as powers of 2, but it still has a fairly traditional "mathematical interpretation": it is the result of iterating

$$n \mapsto \text{Ceiling}[\frac{3n}{2}]$$

or

$$n \mapsto \text{If}[\text{EvenQ}[n], \frac{3n}{2}, \frac{3n+1}{2}]$$

(and this same iteration applies for any initial string of 1s of any length).

But consider now the rule:

{1 → 12, 2 → 111}

Here is what it does starting with sequences of 1s of different lengths:

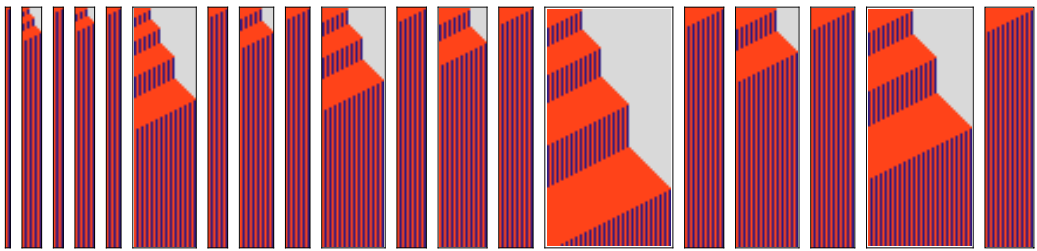

In effect it is taking the initial number of 1s n and computing the function:

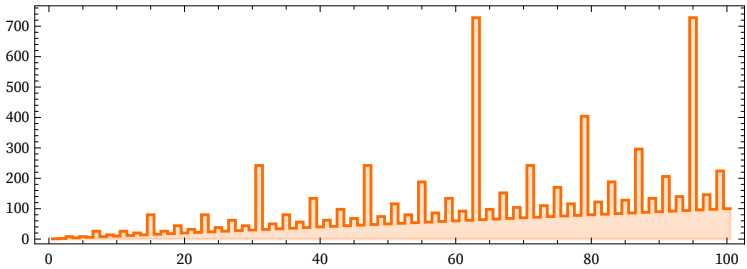

But what "is" this function? In effect it depends on the binary digits of n, and turns out to be given (for n > 1) by:

With[{e = IntegerExponent[n + 1, 2]}, $\frac{3^e (n+1)}{2^e} - 1$]

What other "identifiable functions" can simple tag systems produce? Consider the rules:

{1 → 23, 2 → 1, 3 → 111}

Starting with a string of five 1s this gives (3 is white)

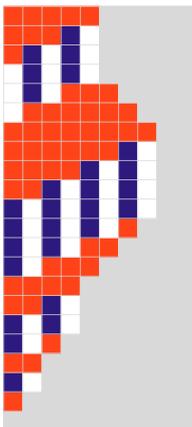

in effect running for 21 steps and then terminating. If one looks at the string of 1s produced here, their sequence of lengths is 5, 8, 4, 2, 1, and in general the sequence is determined by the iteration

n ⟼ If[EvenQ[n], $\frac{n}{2}$, 3 n + 1]

except that if n reaches 1 the tag system terminates, while the iteration keeps going.

So if we ask what this tag system is "doing", we can say it's computing 3n + 1 problem iterations, and we can explicitly "see it doing the computation". Here it's starting with n = 7

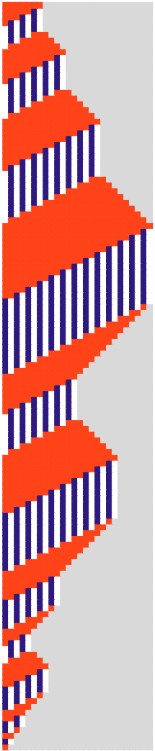

and here it's starting with successive values of n:

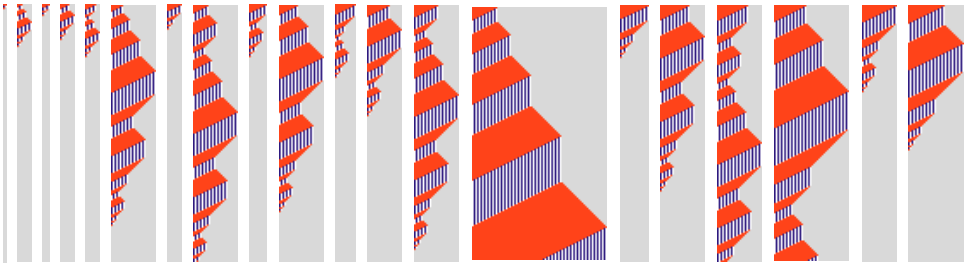

Does the tag system always eventually halt? This is exactly the 3n + 1 problem—which has been unsolved for the better part of a century.

It might seem remarkable that even such a simple tag system rule can in effect give us such a difficult mathematical problem. But the Principle of Computational Equivalence makes this seem much less surprising—and in fact it tells us that we should expect tag systems to quickly "ascend out of" the range of computations to which we can readily assign traditional mathematical interpretations.

Changing the rule to

{1 → 23, 2 → 111, 3 → 1}

yields instead the iteration

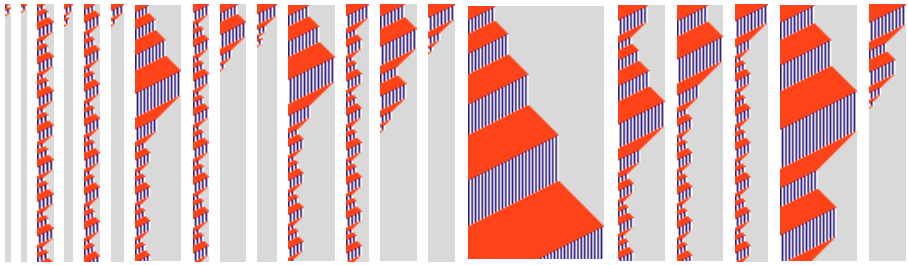

which again is "interpretable" as corresponding to the iteration:

n ⟼ If[EvenQ[n], 3 n/2, (n – 1)/2]

But what if we consider all possible rules, say with the very simple form 1→__, 2→___? Here is what each of the 32 of these does starting from 1111:

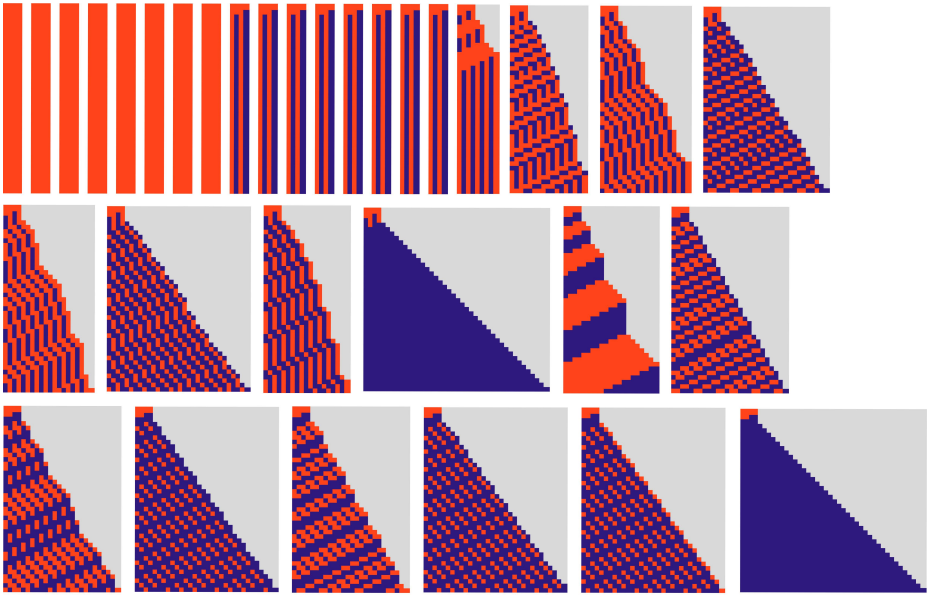

For some of these we've been able to identify "traditional mathematical interpretations", but for many we have not. And if we go even further and look at the very simplest nontrivial rules—of the form 1→_, 2→___—here is what happens starting from a string of 10 1s:

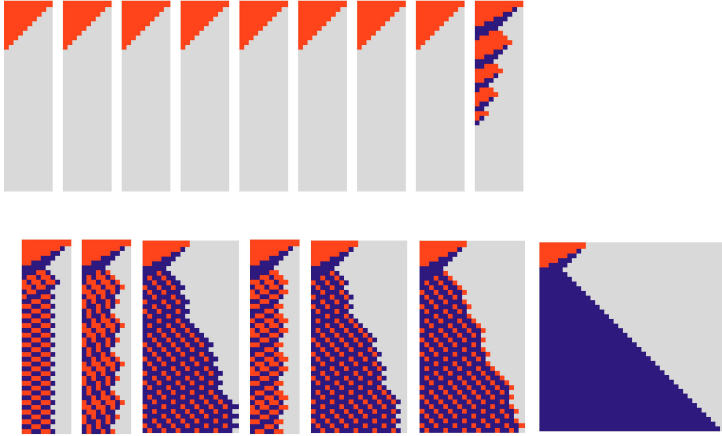

One of these rules we already discussed above:

{1 → 2, 2 → 221}

and we found that it seems to lead to infinite irregular growth (here shown "detrended" by $\sqrt{2}-1$):

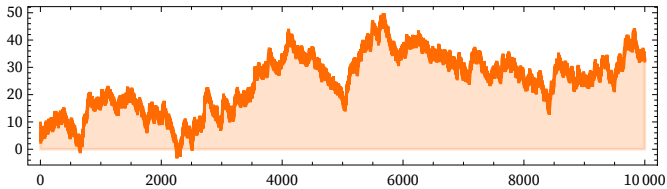

But even in the case of

{1 → 2, 2 → 111}

which appears always to halt

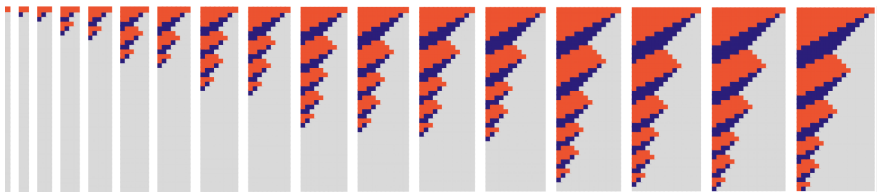

the differences between halting times with successive sizes of initial strings form a surprisingly complex sequence that does not seem to have any simple traditional mathematical interpretation. (By the way, in a case like this it's perfectly possible that there will be some kind of "mathematical interpretation"— though it might be like the page of weird definitions that I found for halting times of Turing machine 600720 in A New Kind of Science.)

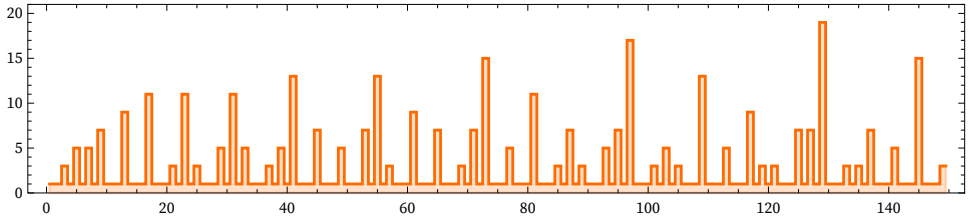

## So Does It Always Halt?

When Emil Post was studying his tag system back in 1921, one of his big questions was: "Does it always halt?" Frustratingly enough, I must report that even a century later I still haven't been able to answer this question.

Running Post's tag system on my computer I'm able to work out what it does billions of times faster than Post could. And I've been able to look at billions of possible initial strings. And I've found that it can take a very long time—like half a trillion steps—for the system to halt:

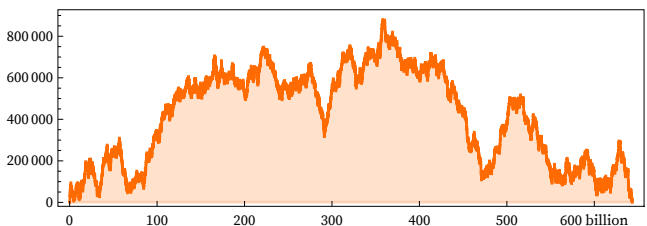

But so far—even with all the computation I've done—I haven't found a single example where it doesn't eventually halt.

If we were doing ordinary natural science, billions of examples that all ultimately work the same would normally be far more than enough to convince us of something. But from studying the computational universe we know that this kind of "scientific inference" won't always be correct. Gödel's theorem from 1931 introduced the idea of undecidability (and it was sharpened by Turing machines, etc.). And that's what can bite us in the computational universe.

Because one of the consequences of undecidability as we now understand it is that there can be questions where there may be no bound on how much computation will be needed to answer them. So this means that even if we have failed to see something in billions of examples that doesn't mean it's impossible; it may just be that we haven't done enough computation to see it.

In practice it's tended to be assumed, though, that undecidability is something rare and exotic, that one will only run into if one asks some kind of awkward—or "meta"—question. But my explorations in the computational universe—and in particular my Principle of Computational Equivalence—imply that this is not correct, and that instead undecidability is quite ubiquitous, and occurs essentially whenever a system can behave in ways that are not obviously simple.

And this means that—despite the simplicity of its construction—it's actually to be expected that something like the 00, 1101 tag system could show undecidability, and so that questions about it could require arbitrary amounts of computational effort to answer. But there's something of a catch. Because the way one normally proves the presence of undecidability is by proving computation universality. But at least in the usual way of thinking about computation universality, a universal system cannot always halt—since otherwise it wouldn't be able to emulate systems that themselves don't halt.

So with this connection between halting and computation universality, we have the conclusion that if the 00, 1101 tag system always halts it cannot be computation universal. So from our failure to find a non-halting example the most obvious conclusion might be that our tag system does in fact always halt, and is not universal.

And this could then be taken as evidence against the Principle of Computational Equivalence, or at least its application to this case. But I believe strongly enough in the Principle of Computational Equivalence that I would tend to draw the opposite conclusion: that actually the 00, 1101 tag system is universal, and won't always halt, and it's just that we haven't gone far enough in investigating it to see a non-halting example yet.

But how far should we have to go? Undecidability says we can't be sure. But we can still potentially use experience from studying other systems to get some sense. And this in fact tends to suggest that we might have to go a long way to get our first non-halting example.

We saw above an example of cellular automata in which unbounded growth (a rough analog of non-halting) does occur, but we have to look through nearly 100,000 initial conditions before we find it. A New Kind of Science contains many other examples. And in number theory, it is quite routine to have Diophantine equations where the smallest solutions are very large.

How should we think about these kinds of things? In essence, we are taking computation universal systems and trying to "program them" (by setting up appropriate initial conditions) to have a particular form of behavior, say non-halting. But there is nothing to say these programs have to be short. Yes, non-halting might seem to us like a simple objective. And, yes, the universal system should in the end be able to achieve it. But given the particular components of the universal system, it may be complicated to get.

Let me offer two analogies. The first has to do with mathematical proofs. Having found the very simplest possible axiom system for Boolean algebra $((p \cdot q) \cdot r) \cdot (p \cdot ((p \cdot r) \cdot p)) == r$, we know that in principle we can prove any theorem in Boolean algebra. But even something like $p \cdot q = q \cdot p$—that might seem simple to us—can take hundreds of elaborate steps to prove given our particular axiom system.

As a more whimsical example, consider the process of self-reproduction. It seems simple enough to describe this objective, yet to achieve it, say with the components of molecular biology, may be complex. And maybe on the early Earth it was only because there were so many molecules, and so much time, that self-reproduction could ever be "discovered".

One might think that, yes, it could be difficult to find something (like a non-halting initial condition, or a configuration with particular behavior in a cellular automaton) by pure search, but that it would still be possible to systematically "engineer" one. And indeed there may be ways to "engineer" initial conditions for the 00, 1101 tag system. But in general it is another consequence of the Principle of Computational Equivalence (and computational irreducibility) that there is no guarantee that there will be any "simple engineering path" to reach any particular capability.

By the way, one impression from looking at tag systems and many other kinds of systems is that as one increases the sizes of initial conditions, one crosses a sequence of thresholds for different behaviors. Only at size 14, for example, might some long "highway" in our tag system's state transition graph appear. And then nothing longer might appear until size 17. Or some particular period of final cycle might only appear at size-15 initial conditions. It's as if there's a "minimum program length" needed to achieve a particular objective, in a particular system. And perhaps similarly there's a minimum initial string length necessary to achieve non-halting in our tag system—that we just don't happen to have reached yet. (I've done random searches in longer initial conditions, though, so we at least know it's not common there.)

OK, but let's try a different tack. Let's ask what would be involved in proving that the tag system doesn't always halt. We're trying to prove essentially the following statement: "There exists an initial condition i such that for all steps t the tag system has not halted". In the language of mathematical logic this is a $\exists\forall$ statement, that is at the $\Sigma_2^0$ level in the [arithmetic hierarchy](.).

One way to prove it is just explicitly to find a string whose evolution doesn't halt. But how would one show that the evolution doesn't halt? It might be obvious: there might for example just be something like a fixed block that is getting added in a simple cycle of some kind, as in:

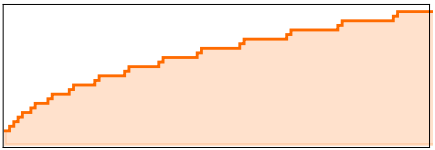

But it also might not be obvious. It could be like some of our examples above where there seems to be systematic growth, but where there are small fluctuations:

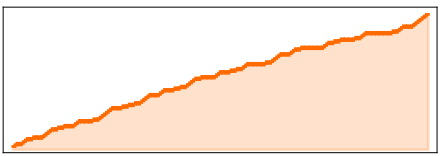

Will these fluctuations suddenly become big and lead the system to halt? Or will they always stay somehow small enough that that cannot happen? There are plenty of questions like this that arise in number theory. And sometimes (as, for example, with the Skewes number associated with the distribution of primes) there can be surprises, with very long-term trends getting reversed only in exceptionally large cases.

By the way, even identifying "halting" can be difficult, especially if (as we do for our tag system) we define "halting" to include going into a cycle. For example, we saw above a tag system that does cycle, but takes more than 18,000 steps to do so:

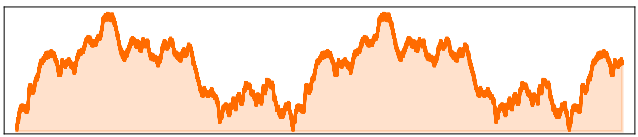

Conversely, just because something takes a long time to halt doesn't mean that it will be difficult to show this. For example, it is quite common to see Turing machines that take a huge number of steps to halt, but behave in basically systematic and predictable ways (this one takes 47,176,870 steps):

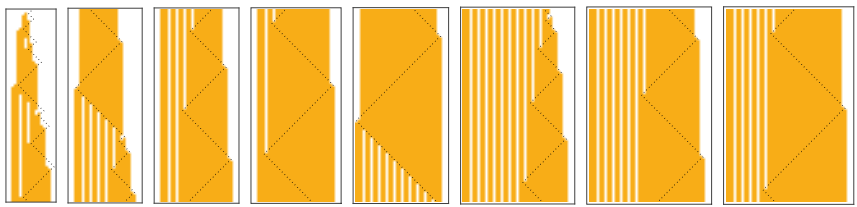

But to "explain why something halts" we might want to have something like a mathematical proof: a sequence of steps consistent with a certain set of axioms that derives the fact that the system halts. In effect the proof is a higher-level ("symbolic") way of representing aspects of what the system is doing. Instead of looking at all the individual values at each step in the evolution of the system we're just calling things x and y (or whatever) and deriving relationships between them at some kind of symbolic level.

And given a particular axiom system it may or may not be possible to construct this kind of symbolic proof of any given fact. It could be that the axiom system just doesn't have the "derivational power" to represent faithfully enough what the system we are studying is doing.

So what does this mean for tag systems? It means, for example, that it could perfectly well be that a given tag system evolution doesn't halt—but that we couldn't prove that using, say, the axiom system of Peano Arithmetic.

And in fact as soon as we have a system that is computation universal it turns out that any finite axiom system must eventually fail to be able to give a finite proof of some fact about the system. We can think of the axioms as defining certain relations about the system. But computational irreducibility implies that eventually the system will be able to do things which cannot be "reduced" by any finite set of relations.

Peano Arithmetic contains as an axiom the statement that mathematical induction works, in the sense that if a statement s[0] is true, and s[n] implies s[n + 1], then any statement s[n] must be true. But it's possible to come up with statements that entail for example nested collections of recursions that effectively grow too quickly for this axiom alone to be able to describe symbolically "in one go" what they can do.

If one uses a stronger axiom system, however, then one will be able to do this. And, for example, Zermelo–Fraenkel set theory—which allows not only ordinary induction but also transfinite induction—may succeed in being able to give a proof even when Peano Arithmetic fails.

But in the end any finitely specified axiom system will fail to be able to prove everything about a computationally irreducible system. Intuitively this is because making proofs is a form of computational reduction, and it is inevitable that this can only go so far. But more formally, one can imagine using a computational system to encode the possible steps that can be made with a given axiom system. Then one would construct a program in the computational system that would systematically enumerate all theorems in the axiom system. (It may be easier to think of first creating a multiway system in which each possible application of the axiom rules is made, and then "unrolling" the multiway system to be "run sequentially".)

And for example we could set things up so that the computational system halts if it ever finds an inconsistency in the theorems derived from the axiom system. But then we know that we won't be able to prove that the computational system does not halt from within the axiom system because (by Gödel's second incompleteness theorem) no nontrivial axiom system can prove its own consistency.

So if we chose to work, say, purely within Peano Arithmetic, then it might be that Post's original question is simply unanswerable. We might have no way to prove or disprove that his tag system always halts. To know that might require a finer level of analysis—or, in effect, a higher degree of reduction—than Peano Arithmetic can provide. (Picking a particular model of Peano Arithmetic would resolve the question, but to home in on a particular model can in effect require infinite computational effort.)

If we have a tag system that we know is universal then it's inevitable that certain things about it will not be provable within Peano Arithmetic, or any other finitely specified axiom system. But for any given property of the system it may be very difficult to determine whether that property is provable within Peano Arithmetic.

The problem is similar to proving computation universality: in effect one has to see how to encode some specified structure within a particular formal system—and that can be arbitrarily difficult to do. So just as it may be very hard to prove that the 00, 1101 tag system is computation universal, it may also be very difficult to prove that some particular property of it is not "accessible" through Peano Arithmetic.

Could it be undecidable whether the 00, 1101 tag system always halts? And if we could prove this, would this actually have proved that it in fact doesn't halt? Recall that above we mentioned that at least the obvious statement of the problem is at the $\Sigma_2^0$ level in the arith-metic hierarchy. And it turns out that statements at this level don't have "default truth values", so proving undecidability wouldn't immediately give us a conclusion. But there's nothing to say that some clever reformulation might not reduce the problem to $\Pi_1^0$ or $\Sigma_1^0$, at which point proving undecidability would lead to a definite conclusion.

(Something like this in fact happened with the Riemann Hypothesis. At first this seemed like a $\Pi_2^0$ statement, but it was reformulated as a $\Pi_1^0$ statement—and eventually reduced to the specific statement several sections above that a particular computation should not terminate. But now if the termination of this is proved undecidable, it must in fact not terminate, and the Riemann Hypothesis must be true.)

Can one prove undecidability without proving computation universality? There are in principle systems that show "intermediate degrees": they exhibit undecidability but cannot directly be used to do universal computation (and Post was in fact the person who suggested that this might be possible). But actual examples of systems with intermediate degree still seem to involve having computation universality "inside", but then limiting the input-output capabilities to prevent the universality from being accessed, beyond making certain properties undecidable.

The most satisfying (and ultimately satisfactory) way to prove universality for the 00, 1101 tag system would simply be to construct a compiler that takes a specification of some other system that is known to support universality (say a particular known-to-be-universal tag system, or the set of all possible tag systems) and then turns this into an initial string for the 00, 1101 tag system. The tag system would then "run" the string, and generate something that could readily be "decoded" as the result of the original computation.

But there are ways one might imagine establishing what amounts to universality, that could be enough to prove halting properties, even though they might not be as "practical" as actual ways to do computations. (Yes, one could conceivably imagine a molecular-scale computer that works just like a tag system.)

In the current proofs of universality for the simplest cellular automata and Turing machines, for example, one assumes that their initial configurations contain "background" periodic patterns, with the specific input for a particular computation being a finite-size perturbation to this background. For a cellular automaton or Turing machine it seems fairly unremarkable to imagine such a background: even though it extends infinitely across the cells of the system it somehow does not seem to be adding more than a small amount of "new information" to the system.

But for a tag system it's more complicated to imagine an infinite periodic "background", because at every step the string the system is dealing with is finite. One could consider modifying the rules of the tag system so that, for example, there is some fixed background that acts as a "mask" every time the block of elements is added at the end of the string. (For example, the mask could flip the value of every $n^{\text{th}}$ element, relative to a fixed "coordinate system".)

But with the original tag system rules the only way to have an infinite background seems to be to have an infinite string. But how could this work? The rules of the tag system add elements at the end of the string, and if the string is infinitely long, it will take an infinite number of steps before the values of these elements ever matter to the actual behavior of the system.

There is one slightly exotic possibility, however, which is to think about transfinite versions of the tag system. Imagine that the string in the tag system has a length given by a transfinite number, say the ordinal $\omega$. Then it is perfectly meaningful in the context of transfinite arithmetic to imagine additional elements being added at positions $\omega + 1$ etc. And if the tag system then runs for $\omega$ steps, its behavior can start to depend on these added elements.

And even though the strings themselves would be infinite, there can still be a finite ("symbolic") way to describe the system. For example, there could be a function f[i] which defines the value of the $i^{th}$ element. Then we can formally write down the rules for the tag system in terms of this function. And even though it would take an infinite time to explicitly generate the strings that are specified, it can still be possible to "reason" about what happens, just by doing symbolic operations on the function f.

Needless to say, the various issues I've discussed above about provability in particular axiom systems may come into play. But there may still be cases where definite results about computation universality could be established "symbolically" about transfinite tag systems. And conceivably such results could then be "projected down" to imply undecidability or other results about tag systems with finite initial strings.

Clearly the question of proving (or disproving) halting for the 00, 1101 tag system is a complicated one. We might be lucky, and be able to find with our computers (or conceivably engineer) an initial string that we can see doesn't halt. Or we might be able to construct a symbolic representation in which we can carry out a proof.

But ultimately we are in a sense at the mercy of the Principle of Computational Equivalence. There is presumably computational irreducibility in the 00, 1101 tag system that we can't systematically outrun.

Yes, the trace of the tag system seems to be a good approximation to a random walk. And, yes, as a random walk it will halt with probability 1. But in reality it's not a "truly random" random walk; it's a walk determined by a specific computational process. We can turn our questions about halting to questions about the randomness of the walk (and to do so may provide interesting connections with the foundations of probability theory). But in the end we're back to the same issues, and we're still confronted by computational irreducibility.

## More about the History

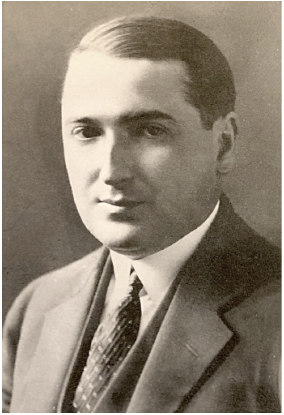

Tag systems are simple enough that it's conceivable they might have arisen in something like games even millennia ago. But for us tag systems—and particularly the specific 00, 1101 tag system we've mostly been studying—were the invention of Emil Post, in 1921.

Emil Post lived most of his life in New York City, though he was born (into a Jewish family) in 1897 in Augustow, Poland (then part of the Russian Empire). (And, yes, it's truly remarkable how many of the notable contributors to mathematical logic in the early part of the 20th century were born to Jewish families in a fairly small region of what's now eastern Poland and western Ukraine.)

As a child, Post seems to have at first wanted to be an astronomer, but having lost his left arm in a freak car-related street accident at age 12 he was told this was impractical—and turned instead to mathematics. Post went to a public high school for gifted students and then attended City College of New York, graduating with a bachelor's degree in math in 1917. Perhaps presaging a lifelong interest in generalization, he wrote his first paper while in college (though it wasn't published until 15+ years later), on the subject of fractional differentiation.

He enrolled in the math PhD program at Columbia, where he got involved in a seminar studying Whitehead and Russell's recently published Principia Mathematica, run by Cassius Keyser, who was one of the early American mathematicians interested in the foundations of math (and who wrote many books on history and philosophy around mathematics; a typical example being his 1922 Mathematical Philosophy, a Study of Fate and Freedom). Early in graduate school, Post wrote a paper about functional equations for the gamma function (related to fractional differentiation), but soon he turned to logic, and his thesis—written in 1920—included early versions of what became his signature ideas.

Post's main objective in his thesis was to simplify, streamline and further formalize Principia Mathematica. He started by looking at propositional calculus, and tried to "drill down" to find out more of what logic was really about. He invented truth tables (as several other people also independently did) and used them to prove completeness and consistency results. He investigated how different logic functions could be built up from one another through composition, classifying different elements of what's now called the Post lattice.

(He commented on NAND and an early simple axiom system for it—and might well have gone further with it if he'd known the minimal axiom system for NAND that I finally discovered in 2000. In another small-intellectual-world story, I realize now his lattice is also similar to my "cellular automaton emulation network".) Going in the direction of "what's logic really about" Post also considered multivalued logic, and algebraic structures around it.

Post published the core of his thesis in 1921 as "Introduction to a General Theory of Elementary Propositions", but—in an unfortunate and recurring theme—didn't publish the whole thing for another 20 years. But even in 1920 Post had what he called "generalization by postulation" and this quickly turned into the idea that all operations in Principia Mathematica (or mathematics in general) could ultimately be represented as transformations ("production rules") on strings of characters.

When he finally ended up publishing this in 1943 he called the resulting formal structures "canonical systems". And already by 1920 he'd discovered that not all possible production rules were needed; it was sufficient to have only ones in "normal form" g\$→\$h, where \$ is a "pattern variable". (The idea of \$ representing a pattern became common in early computer string-manipulation systems, and in fact I used it for expression patterns in my SMP system in 1979—probably without at the time knowing it came from Post.)

Post was close to the concept of universal computation, and the notion that anything (in his case, any string transformation) could be built up from a fixed set of primitives. And in 1920 —in the effort to "reduce his primitives" he came up with tag systems. At the time—11 years before Gödel's theorem—Post and others still thought that it might somehow be possible to "solve mathematics" in some finite way. Post felt he had good evidence that Principia Mathematica could be reduced to string rewriting, so now he just had to solve that.

One basic question was how to tell when two strings should be considered equivalent under the string rewriting rules. And in formulating a simple case of this Post came up with tag systems. In particular, he wanted to determine whether the "iterative process [of tag] was terminating, periodic, or divergent". And Post made "the problem of 'tag'… the major project of [his] tenure of a Procter fellowship in mathematics at Princeton during the academic year 1920–21."

Post later reported that a "major success of the project was the complete solution of the problem for all bases in which $\mu$ and $v$ were both 2", though stated that "even this special case… involved considerable labor". But then, as he later wrote, "while considerable effort was expanded [sic] on the case $\mu = 2$, $v > 2$… little progress resulted… [with] such a simple basis as 0→00, 1→1101, $v = 3$, proving intractable". Post makes a footnote "Numerous initial sequences … tried [always] led… to termination or periodicity, usually the latter." Then he added, reflecting our random walk observations, "It might be noted that an easily derived probability 'prognostication' suggested… that periodicity was to be expected." (I'm curious how he could tell it should be periodicity rather than termination.)

But by the end of the summer of 1921, Post had concluded that "the solution of the general problem of 'tag' appeared hopeless, and with it [his] entire program of the solution of finiteness problems". In other words, the seemingly simple problem of tag had derailed Post's whole program of "solving mathematics".

In 1920 Princeton had a top American mathematics department, and Post went there on a prestigious fellowship (recently endowed by the Procter of Procter & Gamble). But—like the problem of tag—things did not work out so well there for Post, and in 1921 he had the first of what would become a sequence of "runaway mind" manic episodes, in what appears to have been a cycle of what was then called manic depression.

It's strange to think that the problem of tag might have "driven Post crazy", and probably the timing of the onset of manic depression had more to do with his age—though Post later seems to have believed that the excitement of research could trigger manic episodes (which often involved talking intensely about streams of poorly connected ideas, like the "psychic ether" from which new ideas come, discovering a new star named "Post", etc.) But in any case, in late 1921 Post—who had by then returned to Columbia—was institutionalized.

By 1924 he had recovered enough to take up an instructorship at Cornell, but then relapsed. Over the years that followed he supported himself by teaching high school in New York, but continued to have mental health issues. He married in 1929, had a daughter in 1932, and in 1935 finally became a professor at City College, where he remained for the rest of his life.

Post published nothing from the early 1920s until 1936. But in 1936—with Gödel's theorem known, and Alonzo Church's "An Unsolvable Problem of Elementary Number Theory" recently published—Post published a 3-page paper entitled "Finite Combinatory Processes— Formulation 1". Post comes incredibly close to defining Turing machines (he talks about "workers" interacting with a potentially infinite sequence of "marked" and "unmarked boxes"). And he says that he "expects [his] formulation to be logically equivalent to recursiveness in the sense of the Gödel–Church development", adding "Its purpose, however, is not only to present a system of a certain logical potency but also, in its restricted field, of psychological fidelity". Post doesn't get too specific, but he does make the comment (rather resonating with my own work, and particularly our Physics Project) that the hypothesis of global success of these formalisms would be "not so much ... a definition or an axiom but ... a natural law".

In 1936 Post also published his longest-ever paper: 142 pages on what he called "polyadic groups". It's basically about abstract algebra, but in typical Post style, it's a generalization, involving looking not at binary "multiplication" operations but for example ternary ones. It's not been a popular topic, though, curiously, I also independently got interested in it in the 1990s, eventually discovering Post's work on it.

By 1941 Post was publishing more, including several now-classic papers in mathematical logic, covering things like degrees of unsolvability, the unsolvability of the word problem for semigroups, and what's now called the Post Correspondence Problem. He managed his time in a very precise way, following a grueling teaching schedule (with intense and precise lectures planned to the minute) and---apparently to maintain his psychological wellbeing--- restricting his research activities to three specific hours each day (interspersed with walks). But by then he was a respected professor, and logic had become a more popular field, giving him more of an audience.

In 1943, largely summarizing his earlier work, Post published "Formal Reductions of the General Combinatorial Decision Problem", and in it, the "problem of tag" makes its first published appearance:

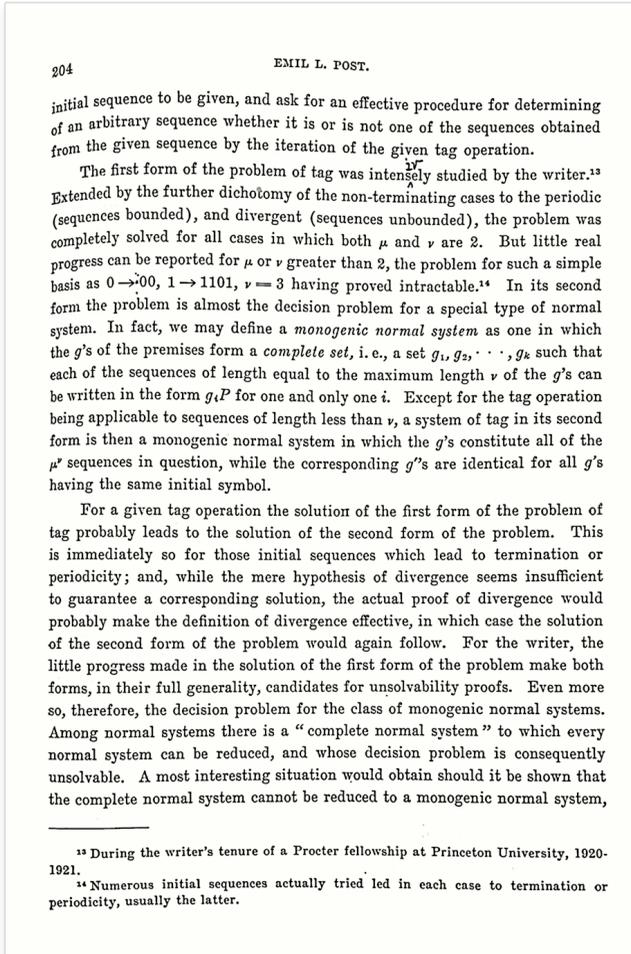

Post notes that "the little progress made in [its] solution" makes it a "candidate for unsolvability". (Notice the correction in Post's handwriting "intensely" → "intensively" in the copy of his paper reproduced in his collected works.)

Through all this, however, Post continued to struggle with mental illness. But by the time he reached the age of 50 in 1947 he began to improve, and even loosened up on his rigid schedule. But in 1954 depression was back, and after receiving electroshock therapy (which he thought had helped him in the past), he died of a heart attack at the age of 57.

His former undergraduate student, Martin Davis, eventually published Post's "Absolutely Undecidable Problems", subtitled "Account of an Anticipation", which describes the arc of Post's work—including more detail on the story of tag systems. And in hindsight we can see how close Post came to discovering Gödel's theorem and inventing the idea of universal computation. If instead of turning away from the complexity he found in tag systems he had embraced and explored it, I suspect he would have discovered not only foundational ideas of the 1930s, but also some of what I found half a century later in my by-then-computer-assisted explorations of the computational universe.

When Post died, he left many unpublished notes. A considerable volume of them concern a major project he launched in 1938 that he planned to call "Creative Logic". He seemed to feel that "extreme abstraction" as a way of exploring mathematics would give way to something in which it's recognized that "processes of deduction are themselves essentially physical and hence subject to formulations in a physical science". And, yes, there's a strange resonance here with my own current efforts—informed by our Physics Project—to "physicalize" metamathematics. And perhaps I'll discover that here too Post anticipated what was to come.

So what happened to tag systems? By the mid-1950s Post's idea of string rewriting ("production systems") was making its way into many things, notably both the development of generative grammars in linguistics, and formal specifications of early computer languages. But tag systems—which Post had mentioned only once in his published works, and then as a kind of aside—were still basically unknown.

Post had come to his string rewriting systems—much as Turing had come to his Turing machines—as a way to idealize the processes of mathematics. But by the 1950s there was increasing interest in using such abstract systems as a way to represent "general computations", as well as brains. And one person drawn in this direction was Marvin Minsky. After a math PhD in 1954 at Princeton on what amounted to analog artificial neural networks, he started exploring more discrete systems, initially finite automata, essentially searching for the simplest elements that would support universal computation (and, he hoped, thinking-like behavior).

Near the end of the 1950s he looked at Turing machines—and in trying to find the simplest form of them that would be universal started looking at their correspondence with Post's string rewriting systems. Marvin Minsky knew Martin Davis from their time together as graduate students at Princeton, and by 1958 Davis was fully launched in mathematical logic, with a recently published book entitled Computability and Unsolvability.

As Davis tells it now, Minsky phoned him about some unsolvability results he had about Post's systems, asking if they were of interest. Davis told him about tag systems, and that Post had thought they might be universal. Minsky found that indeed they were, publishing the result in 1960 in "Recursive Unsolvability of Post's Problem of 'Tag' and Other Topics in Theory of Turing Machines".

Minsky had recently joined the faculty at MIT, but also had a position at MIT's Lincoln Laboratory, where in working on computing for the Air Force there was a collaboration with IBM. And it was probably through this that Minsky met John Cocke, a lifelong computer designer (and general inventor) at IBM (who in later years was instrumental in the development of RISC architecture). The result was that in 1963 Minsky and Cocke published a paper entitled "Universality of Tag Systems with P=2" that dramatically simplified Minsky's construction and showed (essentially by compiling to a Turing machine) that universality could be achieved with tag systems that delete only 2 elements at each step. (One might think of it as an ultimate RISC architecture.)

For several years, Minsky had been trying to find out what the simplest universal Turing machine might be, and in 1962 he used the results Cocke and he had about tag systems to construct a 7-state, 4-color universal machine. That machine remained the record holder for the simplest known universal Turing machine for more than 40 years, though finally now we know the very simplest possible universal machine: a 2,3 machine that I discovered and conjectured would be universal—and that was proved so by Alex Smith in 2007 (thereby winning a prize I offered).

But back in 1967, the visibility of tag systems got a big boost. Minsky wrote an influential book entitled Computation: Finite and Infinite Machines—and the last part of the book was devoted to "Symbol-Manipulation Systems and Computability", with Post's string rewriting systems a centerpiece.

But my favorite part of Minsky's book was always the very last chapter: "Very Simple Bases for Computability". And there on page 267 is Post's tag system:

SEC. 14.5    VERY SIMPLE BASES FOR COMPUTABILITY    267

14.6 THE PROBLEM OF "TAG" AND
     MONOGENIC CANONICAL SYSTEMS

While a graduate student at Princeton in 1921, Post [1965] studied a class of apparently simple but curiously frustrating problems of which the following is an example:

Given a finite string $S$ of 0's and 1's, examine the first letter of $S$. If it is 0, delete the first three letters of $S$ and append 00 to the result. If the first letter is 1, delete the first three letters and append 1101. Perform the same operation on the resulting string, and repeat the process so long as the resulting string has three or more letters.
For example, if the initial string is 10010, we get

```
10010
   101101
      1011101
         1101101
            1101101
               0111011101
                  101110100
                     1011001101
                        100110111101
                           1101110111101
                              1101111011101
                                 011011101110100
                                    11011101001101
                                       111010011011101
                                          0100110111011101
repeats →       011011110110100
                   0111011010000
                      101110100000
                         110100000011101
                            1000000110111101
                               00001101110111101
                          ← 011011101110100
```

The string has grown, but it has just repeated itself (and hence will continue to repeat the last six iterations forever). Suppose that we start with a different string $S'$. The reader might try, for example, $(100)^7$, that is, 100100100100100100100, but he will almost certainly give up without answering the question: "Does this string, too, become repetitive?" In fact, the answer to the more general question "Is there an effective way to decide, for any string $S$, whether this process will ever repeat when started with $S$?" is still unknown. Post found this (00, 1101) problem "intractable," and so did I, even with the help of a computer. Of course, unless one has a theory, one cannot expect much help from a computer

Minsky reports that "Post found this (00, 1101) problem 'intractable', and so did I, even with the help of a computer". But then he adds, in a style very characteristic of the Marvin Minsky I knew for nearly 40 years: "Of course, unless one has a theory, one cannot expect much help from a computer (unless it has a theory)…" He goes on to say that "if the reader tries to study the behavior of 100100100100100100 without [the aid of a computer] he will be sorry".

Well, I guess computers have gotten a lot faster since the early 1960s; for me now it's trivial to determine that this case evolves to a 10-cycle after 47 steps:

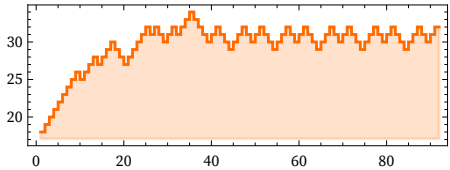

(By the way, I recently asked Martin Davis if Post had ever run a tag system on a computer. He responded: "Goodness! When Post died von Neumann still thought that a dozen computers should suffice for America's needs. I guess I could have programmed [the tag system] for the [Institute for Advanced Study] computer, but it never occurred to me to do so." Notably, in 1954 Davis did start programming logic theorem-proving algorithms on that computer.)

After their appearance in Minsky's book, tag systems became "known", but they hardly became famous, and only a very few papers appeared about them. In 1972, at least their name got some visibility, when Alan Cobham, a longtime IBMer then working on coding theory, published a paper entitled "Uniform Tag Sequences". Yes, this was about tag systems, but now with just one element being deleted at each step, which meant there couldn't really be any interaction between elements. The mathematics was much more tractable (this was one of several inventions of neighbor-independent substitution systems generating purely nested behavior), but it didn't really say anything about Post's "problem of tag".

## Actually, I've Been Here Before…

When I started working on A New Kind of Science in 1991 I wanted to explore the computational universe of simple programs as widely as I could—to find out just how general (or not) the surprising phenomena I'd seen in cellular automata in the 1980s actually were. And almost from the beginning in the table of contents for my chapter on "The World of Simple Programs", nestled between substitution systems and register machines, were tag systems (I had actually first mentioned tag systems in a paper in 1985):

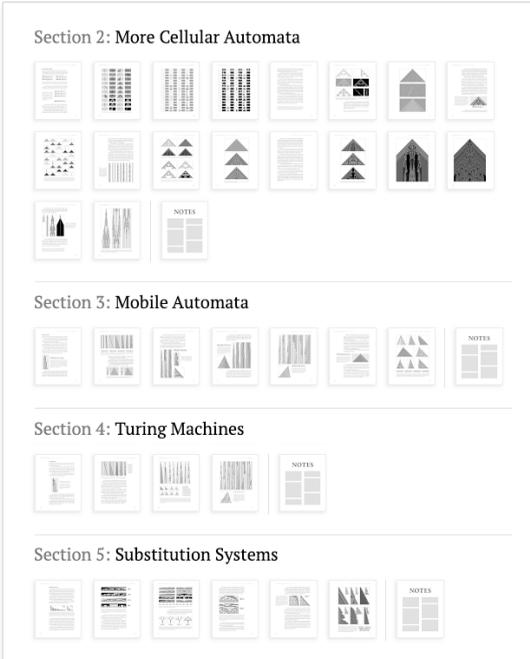
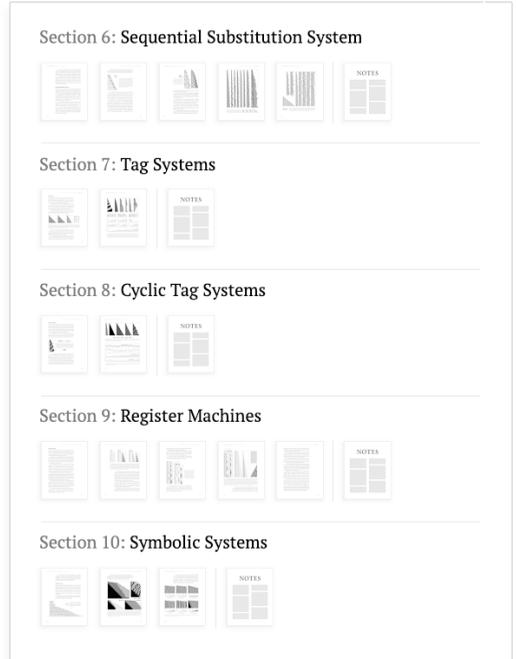

In the main text, I only spent two pages on them:

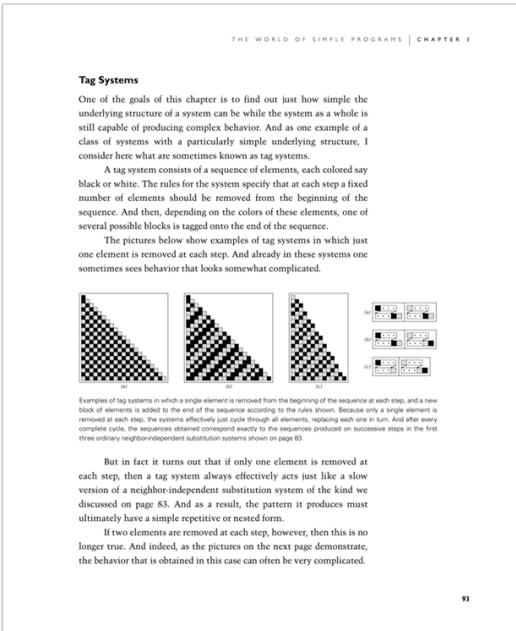
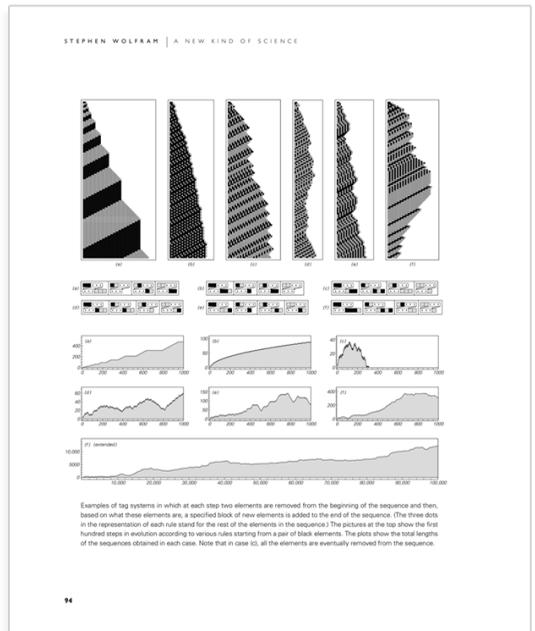

And I did what I have done so many times for so many kinds of systems: I searched and found remarkably simple rules that generate complex behavior. And then on these pages I showed my favorite examples. (I generalized Post's specific tag systems by allowing dependence on more than just the first element.)

Did I look at Post's specific 00, 1101 system? A New Kind of Science includes the note:

> **Tag Systems**
>
> ■ **Implementation.** With the rules for case (a) on page 94 given for example by
> {2, {{0, 0} → {1, 1}, {1, 0} → {}, {0, 1} → {1, 0}, {1, 1} → {0, 0, 0}}}
> the evolution of a tag system can be obtained from
> TSEvolveList[{n_, rule_}, init_, t_] := NestList[If[Length[#] < n, {}, Join[Drop[#, n], Take[#, n] /. rule]] &, init, t]
> An alternative implementation is based on applying to the list at each step rules such as
> {{0, 0, s___} → {s, 1, 1}, {1, 0, s___} → {s},
> {0, 1, s___} → {s, 1, 0}, {1, 1, s___} → {s, 0, 0, 0}}
> There are a total of $((k^{r+1} - 1)/(k - 1))^{k^n}$ possible rules if blocks up to length $r$ can be added at each step and $k$ colors are allowed. For $r = 3$, $k = 2$ and $n = 2$ this is 50,625.
>
> ■ **Page 94 · Randomness.** To get some idea of the randomness of the behavior, one can look at the sequence of first elements produced on successive steps. In case (a), the fraction of black elements fluctuates around 1/2; in (b) it approaches 3/4; in (d) it fluctuates around near 0.3548, while in (e) and (f) it does not appear to stabilize.
>
> ■ **History.** The tag systems that I consider are generalizations of those first discussed by Emil Post in 1920 as simple idealizations of certain syntactic reduction rules in Alfred Whitehead and Bertrand Russell's *Principia Mathematica* (see page 1149). Post's tag systems differ from mine in that his allow the choice of block that is added at each step to depend only on the very first element in the sequence at that step (see however page 670). (The lag systems studied in 1963 by Hao Wang allow dependence on more than just the first element, but remove only the first element.) It turns out that in order to get complex behavior in such systems, one needs either to allow more than two possible colors for each element, or to remove more than two elements from the beginning of the sequence at each step. Around 1921, Post apparently studied all tag systems of his type that involve removal and addition of no more than two elements at each step, and he concluded that none of them produced complicated behavior. But then he looked at rules that remove three elements at each step, and he discovered the rule {3, {{0, _, _} → {0, 0}, {1, _, _} → {1, 1, 0, 1}}}. As he noted, the behavior of this rule varies considerably with the initial conditions used. But at least for all the initial conditions up to length 28, the rule eventually just leads to behavior that repeats with a period of 1, 2, 6, 10, 28 or 40. With more than two colors, one finds that rules of Post's type which remove just two elements at each step can yield complex behavior, even starting from an initial condition such as {0, 0}. An example is {2, {{0, _} → {2, 1}, {1, _} → {0}, {2, _} → {0, 2, 1, 2}}}. (See also pages 1113 and 1141.)

And, yes, it mentions Post's 00, 1101 tag system, then comments that "at least for all the initial conditions up to length 28, the rule eventually just leads to behavior that repeats". An innocuous-looking statement, in very small print, tucked at the back of my very big book. But like so many such statements in the book, there was quite a lot behind it. (By the way, "length 28" then is what I would consider [compressed] length 9 now.)

A quick search of my filesystem quickly reveals (.ma is an earlier format for notebooks that, yes, we can still read over a third of a century later):

| Name | Date Modified | Size | Kind |
|---|---|---|---|
| SSS.nb | Apr 26, 1992 at 12:28 AM | 410 KB | Wolfra...otebook |
| Tag.m | Sep 4, 1991 at 1:09 AM | 524 bytes | Objecti...e code |
| TagAll.ma | Feb 21, 1992 at 12:47 AM | 106 KB | Wolfra...otebook |
| TagAll.nb | Dec 9, 1996 at 2:29 AM | 126 KB | Wolfra...otebook |
| TagNew.ma | Mar 22, 1992 at 9:38 PM | 441 KB | Wolfra...otebook |
| TagNew.nb | Mar 22, 1992 at 9:38 PM | 1.5 MB | Wolfra...otebook |
| TagSystems-new.ma | Mar 25, 1992 at 7:32 PM | 3.4 MB | Wolfra...otebook |
| TagSystems-new.nb | Mar 25, 1992 at 7:32 PM | 9.9 MB | Wolfra...otebook |
| TagSystems.ma | Sep 3, 1991 at 4:33 PM | 607 KB | Wolfra...otebook |
| TagSystems.nb | Dec 9, 1996 at 2:31 AM | 635 KB | Wolfra...otebook |
| TagSystems2.ma | Sep 4, 1991 at 12:39 PM | 259 KB | Wolfra...otebook |
| TagSystems2.nb | Dec 9, 1996 at 2:31 AM | 275 KB | Wolfra...otebook |
| tagtime | Sep 5, 1991 at 12:32 PM | 107 KB | Unix executable |
| tagtime.tm.c | Sep 5, 1991 at 12:32 PM | 2 KB | C Source |
| TagTimes.ma | Sep 6, 1991 at 12:22 AM | 3.2 MB | Wolfra...otebook |
| TagTimes.nb | Sep 6, 1991 at 12:22 AM | 3.4 MB | Wolfra...otebook |
| TM-3-2.ma | May 28, 1994 at 12:37 AM | 204 KB | Wolfra...otebook |

I open one of the notebook files (and, yes, windows—and screens—were tiny in those days):

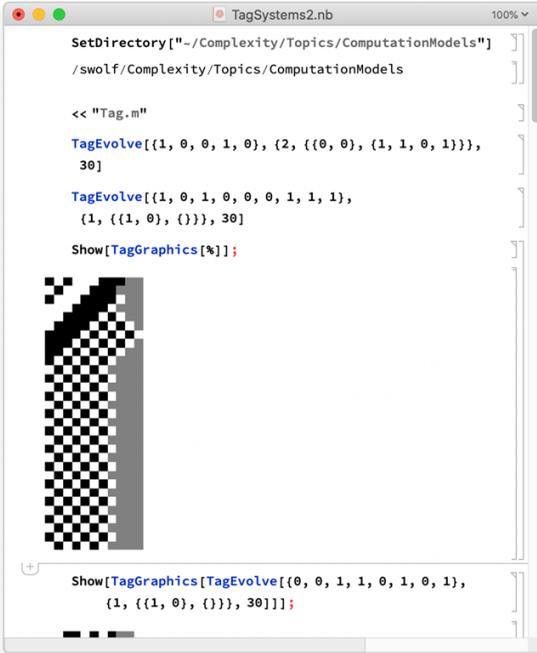

And there it is! Post's 00, 1101 tag system, along with many others I was studying. And it seems I couldn't let go of this; in 1994 I was running a standalone program to try to find infinitely growing cases. Here's the output:

```
pyrethrum [55] 1994> PostTagStand
ilen = 2
p=2: {1, 2}
ilen = 3
ilen = 4
ilen = 5
p=6: {9, 5}
ilen = 6
ilen = 7
ilen = 8
ilen = 9
ilen = 10
ilen = 11
ilen = 12
ilen = 13
ilen = 14
ilen = 15
p=10: {585, 15}
ilen = 16
ilen = 17
ilen = 18
p=28: {4169, 18}
ilen = 19
ilen = 20
ilen = 21
ilen = 22
ilen = 23
ilen = 24
ilen = 25
ilen = 26
p=40: {17076809, 26}
ilen = 27
survived 10000: {2134601, 27}
survived 10000: {2134603, 27}
survived 10000: {2134605, 27}
survived 10000: {2134607, 27}
survived 10000: {2134617, 27}
survived 10000: {2134619, 27}
<many screens>
ilen = 28
```

So that's where I got my statement about "up to size 28" (now size 9) from. I don't know how long this took to run; "pyrethrum" was at the time the fastest computer at our company—with a newfangled 64-bit CPU (a DEC Alpha) running at the now-snail-sounding clock speed of 150 MHz.

My archives from the early 1990s record a fair amount of additional "traffic" about tag systems. Interactions with Marvin Minsky. Interactions with my then-research-assistant about what I ended up calling "cyclic tag systems" (I originally called them "cyclic substitution systems").

For nearly 15 years there's not much. That is, until June 25, 2007. It's been my tradition since we started our Wolfram Summer School back in 2003 that on the first day I do a "live experiment", and try to discover something. Well, that day I decided to look at tag systems. Here's how I began:

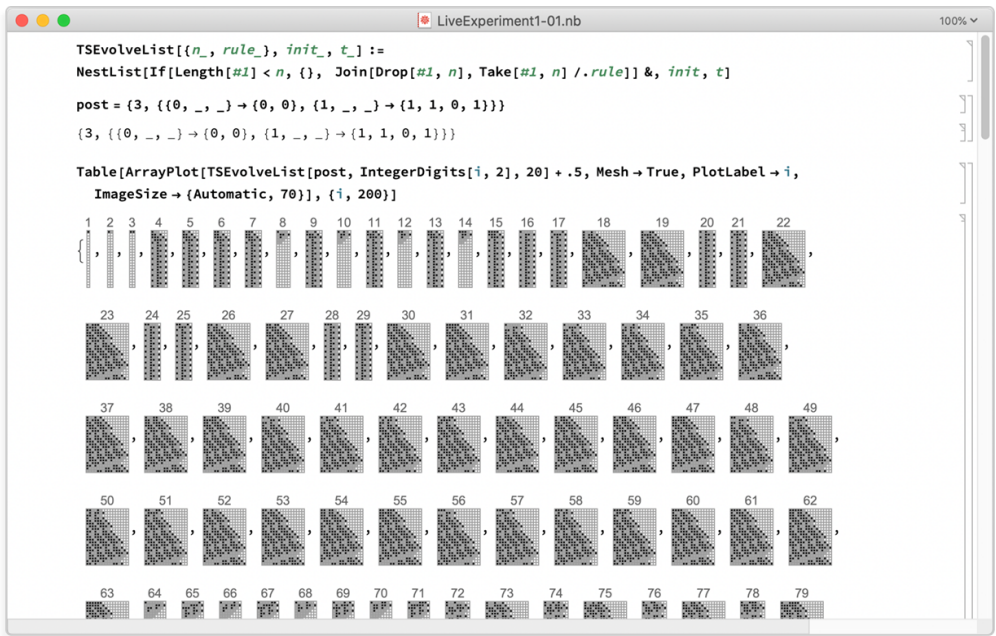

Right there, it's Post's 00, 1101 system. And I think I took it further than I'd ever done before. Pretty soon I was finding "long survivors" (I even got one that lasted more than 200,000 steps):

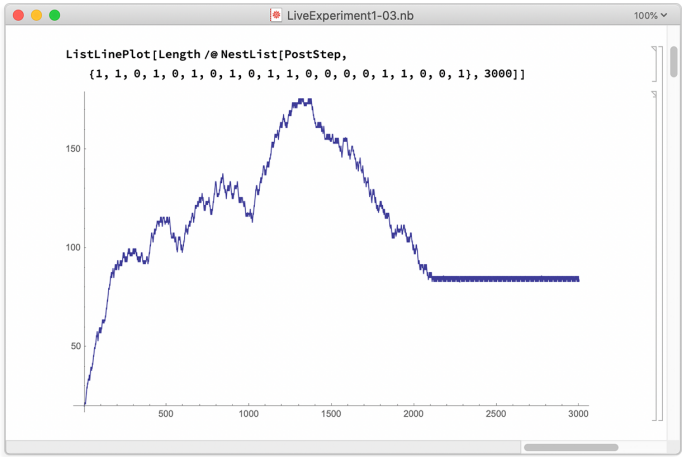

I was drawing state transition graphs:

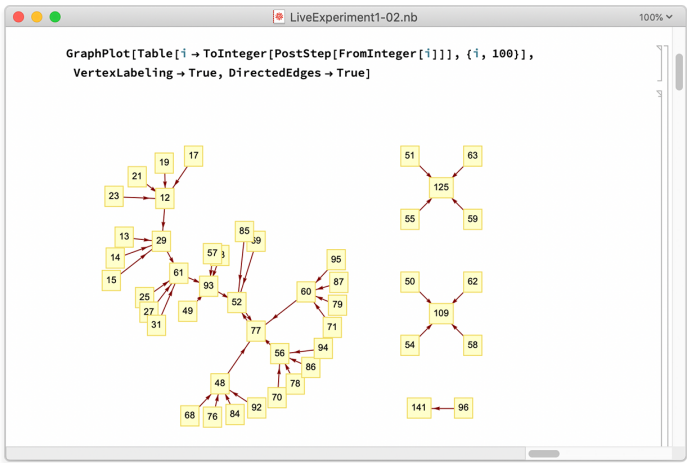

But I obviously decided that I couldn't get further with the 00, 1101 system that day. So I turned to "variants" and quickly found the 2-element-deletion 1, 110 rule that I've described above.

I happened to write a piece about this particular live experiment ("Science: Live and in Public"), and right then I made a mental note: let me look at Post's tag system again before its centenary, in 2021. So here we are....

## The Path Forward

Emil Post didn't manage to crack his 00, 1101 tag system back in 1921 with hand calculations. But we might imagine that a century later—with the equivalent of tens of billions times more computational power we'd be able to do. But so far I haven't managed it.

For Post, the failure to crack his system derailed his whole intellectual worldview. For me now, the failure to crack Post's system in a sense just bolsters my worldview—providing yet more indication of the strength and ubiquity of computational irreducibility and the Principle of Computational Equivalence.

After spending several weeks throwing hundreds of modern computers and all sorts of computational methods at Post's 00, 1101 tag system, what do we know? Here's a summary:

- All $2^{84}$ initial strings up to (uncompressed) length 84 lead either to cycles or termination
- The time to termination or cycling can be as long as 643 billion steps
- The sequence of lengths of strings generated seems to always behave much like a random walk
- The sequences of 0s and 1s generated seem effectively random, apart from about 31% statistical redundancy
- Most cycles are in definite families, but there are also some sporadic ones

What's missing here? Post wanted to know whether the system would halt, and so do we. But now the Principle of Computational Equivalence makes a definite prediction. It predicts that the system should be capable of universal computation. And this basically has the implication that the system can't always halt: there has to be some initial string that will make it grow forever.

In natural science it's standard for theories to make predictions that can be investigated by doing experiments in the physical world. But the kind of predictions that the Principle of Computational Equivalence makes are more general; they're not just about particular systems in the natural world, but about all possible abstract systems, and in a sense all conceivable universes. But it's still possible to do experiments about them, though the experiments are now not physical ones, but abstract ones, carried out in the computational universe of possible programs.

And with Post's tag system we have an example of one particular such experiment: can we find non-halting behavior that will validate the prediction that the system can support universal computation? To do so would be another piece of evidence for the breadth of applicability of the Principle of Computational Equivalence.

But what's going to be involved in doing it? Computational irreducibility tells us that we can't know.

Traditional mathematical science has tended to make the assumption that once you know an abstract theory for something, then you can work out anything you want about it. But computational irreducibility shows that isn't true. And in fact it shows how there are fundamental limitations to science that intrinsically arise from within science itself. And our difficulty in analyzing Post's tag system is in a sense just an "in your face" example of how strong these limitations can be.

But the Principle of Computational Equivalence says that somewhere we'll see non-halting behavior. It doesn't tell us exactly what that behavior will be like, or how difficult it'll be for us to interpret what we see. But it says that the "simple conclusion" of "always halting" shouldn't continue forever.

I've so far done nearly a quintillion iterations of Post's tag system in all. But that hasn't been enough. I've been able to optimize the computations a bit. But fundamentally I've been left with what seems to be raw computational irreducibility. And to make progress I seem to need more time and more computers.

Will a million of today's computers be enough? Will it take a billion? I don't know. Maybe it requires a new level of computational speed. Maybe to resolve the question requires more steps of computation than the physical universe has ever done. I don't know for sure. But I'm optimistic that it's within the current computational capabilities of the world to find that little string of bits for the tag system that will allow us to see more about the general Principle of Computational Equivalence and what it predicts.

In the future there will be ever more that we will want and need to explore in the computational universe. And in a sense the problem of tag is a dry run for the kinds of things that we will see more and more often. But with the distinction of a century of history it's a good place to rally our efforts and learn more about what's involved.

So far it's only been my computers that have been working on this. But we'll be setting things up so that anyone can join the project. I don't know if it'll get solved in a month, a year or a century. But with the Principle of Computational Equivalence as my guide I'm confident there's something interesting to discover. And a century after Emil Post defined the problem I, for one, want to see it resolved.

## Notes

The main tag-system-related functions used are in the Wolfram Function Repository, as TagSystemEvolve, TagSystemEvolveList, TagSystemConvert, CyclicTagSystemEvolveList.

A list of t steps in the evolution of the tag system from an (uncompressed) initial list *init* can be achieved with

TagSystemEvolveList[init_List, t_Integer] := With[
    {ru = Dispatch[{{0, _, _, s___} → {s, 0, 0}, {1, _, _, s___} → {s, 1, 1, 0, 1}}]}, NestList[Replace[ru], init, t]]

or

TagSystemEvolveList[init_List, t_Integer] :=
    NestWhileList[Join[Drop[#, 3], {{0, 0}, {1, 1, 0, 1}}[[1 + First[#]]]] &, init, Length[#] ≥ 3 &, 1, t]

giving for example:

TagSystemEvolveList[{1, 0, 0, 1, 0}, 4]

{{1, 0, 0, 1, 0}, {1, 0, 1, 1, 0, 1}, {1, 0, 1, 1, 1, 0, 1}, {1, 1, 0, 1, 1, 1, 0, 1}, {1, 1, 1, 0, 1, 1, 1, 0, 1}}

The list of lengths can be obtained from

```
TagSystemLengthList[init_List, t_Integer] :=
    Reap[NestWhile[(Sow[Length[#]]; #) &[Join[Drop[#, 3], {{0, 0}, {1, 1, 0, 1}}[[1 + First[#]]]]] &,
        init, Length[#] ≥ 3 &, 1, t]][[2, 1]]
```

giving for example:

```
TagSystemLengthList[{1, 0, 0, 1, 0, 0, 1, 0, 0, 0, 0, 0}, 25]

{13, 14, 15, 14, 15, 16, 15, 16, 15, 16, 17, 18, 19, 20, 21, 22, 21, 22, 23, 24, 23, 24, 25, 26, 27}
```

The output from t steps of evolution can be obtained from:

```
TagSystemEvolve[init_List, t_Integer] :=
    NestWhile[Join[Drop[#, 3], {{0, 0}, {1, 1, 0, 1}}[[1 + First[#]]]] &, init, Length[#] ≥ 3 &, 1, t]
```

A version of this using a low-level queue data structure is:

```
TagSystemEvolve[init_List, t_Integer] :=
    Module[{q = CreateDataStructure["Queue"]}, Scan[q["Push", #] &, init]; Do[If[q["Length"] ≥ 3,
        Scan[q["Push", #] &, If[q["Pop"] == 0, {0, 0}, {1, 1, 0, 1}]]]; Do[q["Pop"], 2]], t]; Normal[q]]
```

The compressed {*p*, *values*} form of a tag system state can be obtained with

```
TagSystemCompress[list_] := {Mod[Length[list], 3], Take[list, 1 ;; -1 ;; 3]}
```

while an uncompressed form can be recovered with

```
TagSystemUncompress[{p_, list_}, pad_ : 0] :=
    Join[Riffle[list, Splice[{pad, pad}]], Table[pad, <|0 → 2, 1 → 0, 2 → 1|>[p]]]
```

Each step in evolution in compressed form is obtained from

```
TagSystemCompressedStep [{p_, {s_, r___}}] := Apply[{#1, Join[{r}, #2]} &,
    <|{0, 0} → {2, {0}}, {1, 0} → {0, {}}, {2, 0} → {1, {0}},
        {0, 1} → {1, {1, 1}}, {1, 1} → {2, {0}}, {2, 1} → {0, {1}}|>[{p, s}]]
```

or:

```
TagSystemCompressedStep [list : {_Integer, _List}] :=
    Replace[list, {{0, {0, s___}} → {2, {s, 0}}, {1, {0, s___}} → {0, {s}}, {2, {0, s___}} → {1, {s, 0}},
        {0, {1, s___}} → {1, {s, 1, 1}}, {1, {1, s___}} → {2, {s, 0}}, {2, {1, s___}} → {0, {s, 1}}}]
```

The largest-scale computations done here made use of further-optimized code (available in the Wolfram Function Repository), in which the state of the tag system is stored in a bit-packed array, with 8 updates being done at a time by having a table of results for all 256 cases and using the first byte of the bit-packed array to index into this. This approach routinely achieves a quarter billion updates per second on current hardware. (Larger update tables no longer fit in L1 cache and so typically do not help.)

As I've mentioned, there isn't a particularly large literature on the specific behavior of tag systems. In 1963 Shigeru Watanabe described the basic families of cycles for Post's 00, 1101 tag system (though did not discover the "sporadic cases"). After A New Kind of Science in 2002, I'm aware of one extensive series of papers (partly using computer experiment methods) written by Liesbeth De Mol following her 2007 PhD thesis. Carlos Martin (a student at the Wolfram Summer School) also wrote about probabilistic methods for predicting tag system evolution.

## Thanks, etc.

Thanks to Max Piskunov and Mano Namuduri for help with tag system implementations, Ed Pegg for tag system analysis (and for joining me in some tag system "hunting expeditions"), Matthew Szudzik and Jonathan Gorard for clarifying metamathematical issues, and Catherine Wolfram for help on the theory of random walks. Thanks also to Martin Davis and Margaret Minsky for clarifying some historical issues (and Dana Scott for having also done so long ago).

## You Can Help!

We're in the process of setting up a distributed computing project to try to answer Emil Post's 100-year old tag system question. Let us know if you'd like to get involved….

## References

*Links to references are included within the body of this document.*